\UseRawInputEncoding
\documentclass[journal,twocolumn]{IEEEtran}
\usepackage{url}
\usepackage{cite}
\usepackage{fancyhdr}
\usepackage{float}
\usepackage{graphicx}
\usepackage{epsfig}
\usepackage{bm}
\usepackage{amsmath}
\usepackage{mathrsfs}
\usepackage{subfigure}
\usepackage{array}
\usepackage{booktabs}
\usepackage{amssymb}
\usepackage{multicol}
\usepackage{multirow}
\usepackage{graphicx}
\usepackage{epstopdf}
\usepackage{algorithm}
\usepackage{amsthm}
\usepackage{bbm}
\usepackage{graphicx}
\usepackage{epsfig}
\usepackage{color}
\usepackage{cite}

\newtheorem{defn}{Definition}

\makeatother

\usepackage{algpseudocode}
\usepackage{amsmath}

\title{Trusted AI in Multi-agent Systems: An Overview of Privacy and Security for Distributed Learning}

\begin{document}
\author{\IEEEauthorblockN{
Chuan Ma, \emph{Member, IEEE},
Jun Li, \emph{Senior Member, IEEE},
Kang Wei, \emph{Member, IEEE},\\
Bo Liu, \emph{Senior Member, IEEE},
Ming Ding, \emph{Senior Member, IEEE},
Long Yuan,\\
Zhu Han, \emph{Fellow, IEEE},
and H. Vincent Poor,~\IEEEmembership{Life~Fellow,~IEEE}
\thanks{This work was supported in part by the National Key RD Program of China 2022YFF0712100, in part by the National Natural Science Foundation of China under Grant 62002170 and 61902184, in part by the Fundamental Research Funds for the Central Universities with No. 30921013104, in part by the Science and Technology on Information Systems Engineering Laboratory WDZC20205250411in part by the Future Network Grant of Provincial Education Board in Jiangsu, in part by the Youth Foundation Project (No. K2023PD0AA01) of Zhejiang Lab, in part by the Research Initiation Project of Zhejiang Lab, and in part by the U.S National Science Foundation under Grants CNS-2128448 and ECCS-2335876. (Corresponding authors: Jun Li and Kang Wei)}
\thanks{Chuan Ma is with Zhejiang Laboratory, Hangzhou 311121. He is also with Key Laboratory of Computer Network and Information Integration (Southeast University), Ministry of Education, China. (Email: chuan.ma@zhejianglab.edu.cn) }
\thanks{Jun Li and Kang Wei are with the School of Electronic and Optical Engineering, Nanjing University of Science and Technology, Nanjing 210096, China. Kang Wei is also with the Department of Computing, Hong Kong Polytechnic University, Hong Kong 999077, China (Email: \{jun.li, kang.wei\}@njust.edu.cn).}
\thanks{Bo Liu is with the University of Technology Sydney, NSW 2007, Australia (Email: bo.liu@uts.edu.au).}
\thanks{Ming Ding is with Data61, CSIRO, NSW 2015, Australia (Email: Ming.Ding@data61.csiro.au).}
\thanks{Long Yuan is with the School of Computer Science, Nanjing University of Science and Technology, Nanjing 210096, China (Email: longyuan@njust.edu.cn).}
\thanks{Zhu Han is with the Department of Electrical and Computer Engineering in the University of Houston, Houston, TX 77004 USA, and also with the Department of Computer Science and Engineering, Kyung Hee University, Seoul, South Korea (Email: zhan2@uh.edu).}
\thanks{H. Vincent Poor is with the Department of Electrical and Computer Engineering, Princeton University, Princeton, NJ 08544 USA (Email: poor@princeton.edu).}
}}

\maketitle
\begin{abstract}
Motivated by the advancing computational capacity of distributed end-user equipment (UE), as well as the increasing concerns about sharing private data, there has been considerable recent interest in machine learning (ML) and artificial intelligence (AI) that can be processed on distributed UEs. Specifically, in this paradigm, parts of a ML process are outsourced to multiple distributed UEs. Then the processed information is aggregated on a certain level at a central server, which turns a centralized ML process into a distributed one, and brings about significant benefits. However, this new distributed ML paradigm raises new risks of privacy and security issues. In this paper, we provide a survey on the emerging security and privacy risks of distributed ML from a unique perspective of information exchange levels, which are defined according to the key steps of a ML process, i.e.: i) the level of pre-processed data, ii) the level of learning models, iii) the level of extracted knowledge and, iv) the level of intermediate results. We explore and analyze the potential of threats for each information exchange level based on an overview of the current state-of-the-art attack mechanisms, and then discuss the possible defense methods against such threats. Finally, we complete the survey by providing an outlook on the challenges and possible directions for future research in this critical area.
\end{abstract}
\section{Introduction}
An explosive growth in data availability arising from proliferating Internet of Things (IoT) and 5G/6G technologies, combined with the availability of increasing computational resources through cloud and data servers, promote the applications of machine learning (ML) in many domains (e.g., finance, health-care, industry, and smart city). ML technologies, e.g., deep learning, have revolutionized the ways that information is extracted with ground-breaking successes in various areas~\cite{8763885}. Meanwhile, owing to the advent of IoT, the number of intelligent applications with edge computing, such as smart manufacturing, intelligent transportation, and intelligent logistics, is growing dramatically.

\begin{figure}
    \centering
    \includegraphics[width=0.45\textwidth]{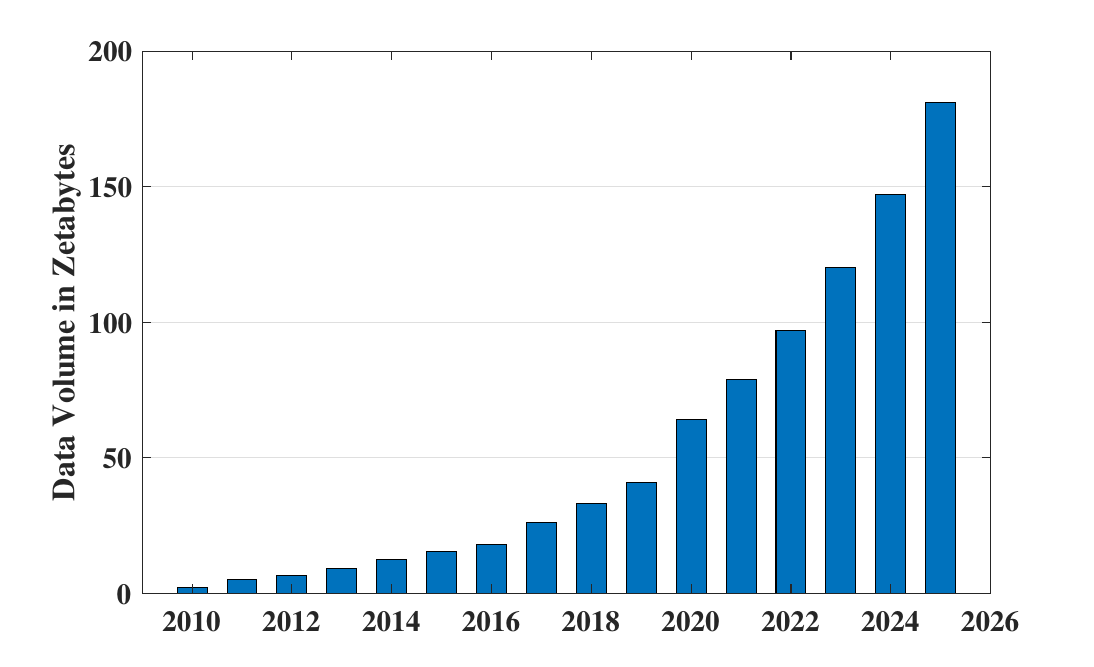}
    \caption{Volume of data/information created, captured, copied, and consumed worldwide from 2010 to 2025.}
    \label{data_inc}
\end{figure}
As such, conventional centralized deep learning can no longer efficiently process the dramatically increased amount of data from the massive number of IoT or edge devices. For example, as shown in Fig.~\ref{data_inc}, the expected volume of data will be 181 zetabytes in 2025\footnote{https://www.statista.com/statistics/871513/worldwide-data-created/}. In addition, the long runtime of training models steers solution designers towards using distributed systems for an increase of parallelization and the total amount of wireless bandwidth, as the training data required for sophisticated applications can easily be on the order of terabytes~\cite{elsebakhi2015large}.
Examples include transaction processing for larger enterprises on data that is stored in different locations~\cite{Bernstein2009Principles} or astronomical data that is too large to move and collect~\cite{raicu2006astroportal}.

To address this challenge, distributed learning frameworks have emerged.
A typical distributed learning fashion involves the cooperation of multiple clients and servers, which thus involves a decentralization and aggregation process along with the ML process \cite{Gu2019Distributed}. With the increasing capability of edge devices, distributed clients are able to execute simple ML tasks. For example, federated learning (FL)~\cite{Jakub2016Federated,Tian2020Federated,Chuan2020On} enables the decoupling of data provisioning by distributed clients and aggregating ML models at a centralized server. In certain ML tasks, the model sometimes can be so large that it cannot be trained in a reasonable amount of time, and cannot run completely on a single machine. Therefore, large-scale distributed ML is proposed in~\cite{yang2019federated} where datasets in each client will be re-analyzed and pre-trained locally, and the knowledge is aggregated by a central server. In addition, aggregating learning results~\cite{papernot2016semi} by the server is another part of distributed ML technology.

To complete a ML task successfully, we need to preserve the integrity and security of the system, along with the privacy of participating clients. As the manufacturers can potentially fail to implement a robust security system in distributed devices, experts on security have warned of potential risks of large numbers of unsecured devices connecting to the Internet~\cite{Zhang2014IoT}. Security and privacy are very significant issues for distributed ML, which introduce a new level of emergent concerns for participants. This is because these devices collect not only personal and sensitive information, e.g., names and telephone numbers but also monitor daily activities. Due to the regular stream of news stories about privacy leakage through major data breaches, users are wary of using personal data in public or private ML tasks with good reasons~\cite{Yang2017A}.

There are some related surveys on security and privacy issues in distributed ML. For example, the challenges and opportunities of distributed learning over conventional (centralized) ML were discussed in \cite{peteiro2013survey,Verbraeken2020A}, which elaborated on limited privacy and security issues. In~\cite{de2020overview,Liu2021When}, the authors focused on the adversarial models related to private information leakage and corresponding defensive mechanisms in ML, and the work~\cite{vepakomma2018no} investigated privacy issues in distributed ML. Moreover, differential privacy (DP) based protection methods were introduced in~\cite{gong2020survey}. In addition, to protect the privacy of the IoT data, the work \cite{Mohammad2020A} surveyed the ML-based method to address the privacy issues including scalability, inter-operability, and limitations on resources, such as computation and energy. The works~\cite{briggs2020review,Enthoven2021An,lyu2020threats} addressed security and privacy issues in FL, together with related solutions. The summary of the related surveys on security and privacy issues in ML is listed in Table~\ref{survey}.

\begin{table*}
\newcommand{\tabincell}[2]{\begin{tabular}{@{}#1@{}}#2\end{tabular}}
\centering
\caption{Existing surveys on private and secure machine learning}\label{survey}
\begin{tabular}{|c|c|l|}
  \hline
  \textbf{Related Survey}&\textbf{Topic}&\textbf{Key contributions}\\
  \hline
  \cite{briggs2020review} &
  \tabincell{l} {Privacy preserving in federated learning \\for IoT data}&
  \tabincell{l} {This work mainly focused on the survey on the use of federated learning for private \\data analysis in IoT, i.e., highly skewed non-IID data with high temporal variability, \\to address privacy concerns, bandwidth limitations and power/compute limitations.} \\
  \hline
  \cite{Mohammad2020A}&
  \tabincell{l} {Machine learning-based solutions to protect \\privacy in IoT}&
  \tabincell{l}{This work surveyed the works that leverage machine learning as a strategy to address \\the privacy issues of IoT including scalability, inter-operability, and resource limitation \\such as computation and energy.}\\
  \hline
  \cite{Xu2019Data}&
  \tabincell{l}{Data Security Issues in Deep Learning}&
  \tabincell{l}{This survey investigated the potential threats of deep learning with respect to black \\and white box attacks and presented related countermeasures on offense and defense.}\\
  \hline

  \cite{gong2020survey}&
  \tabincell{l}{Differentially private machine learning}&
  \tabincell{l}{This survey investigated the existing differentially private machine learning technologies \\and categorized them as the Laplace/Gaussian/exponential mechanism and the output\\ /objective perturbation mechanism}\\
  \hline
  \cite{peteiro2013survey,Verbraeken2020A}&
  \tabincell{l}{Machine learning in distributed systems}&
  \tabincell{l}{These articles provided an overview by outlining the challenges and opportunities \\of distributed machine learning over conventional (centralized) machine learning,  \\and discussing available techniques.}\\
  \hline
  \cite{Enthoven2021An,lyu2020threats,briggs2020review}&
  \tabincell{l}{Attacks and defensive strategies on federated \\deep learning}&
  \tabincell{l}{These works investigated existing vulnerabilities of FL and subsequently provided a \\literature study of defensive strategies and algorithms for FL aimed to overcome \\these attacks.}\\
  \hline
  \cite{de2020overview,Liu2021When,xu2021privacypreserving}&
  \tabincell{l}{Privacy in machine learning}&
  \tabincell{l}{These surveys focused on machine learning and algorithms related to private information \\leakage and corresponding defensive mechanisms.}\\
  \hline
  \cite{vepakomma2018no}&
  \tabincell{l}{Privacy in distributed machine learning}&
  \tabincell{l}{This work focused on the privacy leakage issues in distributed learning and studied \\benefits, limitations, and trade-offs for defensive algorithms.}\\
  \hline
  Our paper&
  \tabincell{l}{Privacy and Security in distributed learning}&
  \tabincell{l}{Our work is different from the above survey articles in the following aspects: 1. our \\work first develops a distributed framework into four levels; 2. the state-of-the-art \\on the private and secure issues in each level are investigated and summarized; \\3. the characteristics of the adversary at each level are further discussed. }\\
  \hline
\end{tabular}
\label{table1}
\end{table*}

Different from the above-mentioned surveys, in this work,
\begin{itemize}
  \item We first give a clear and fresh definition of distributed learning, and develop the distributed learning framework in four levels in terms of sharing different information, namely sharing data, sharing model, sharing knowledge, and sharing results.
  \item We then provide an extensive overview of the current state-of-the-art related to the attacks and defensive mechanisms on the privacy and security issues for each level. Real examples are also listed for each level.
  \item In addition, learned lessons from each aspect are described, which can indeed help readers to avoid potential mistakes.
  \item Several research challenges and future directions are further discussed, which can provide insights into the design of advanced learning paradigms.
\end{itemize}

\section{Background of Distributed ML and the Paper Structure}
In Section~\uppercase\expandafter{\romannumeral2}, we first describe the detailed process that how a machine learning task is executed, and then transit the centralized learning to distributed paradigms, and develop a decentralized learning framework. In addition, we provide descriptions of several widely-studied distributed learning frameworks.
\begin{figure*}
    \centering
    \includegraphics[width=0.95\textwidth]{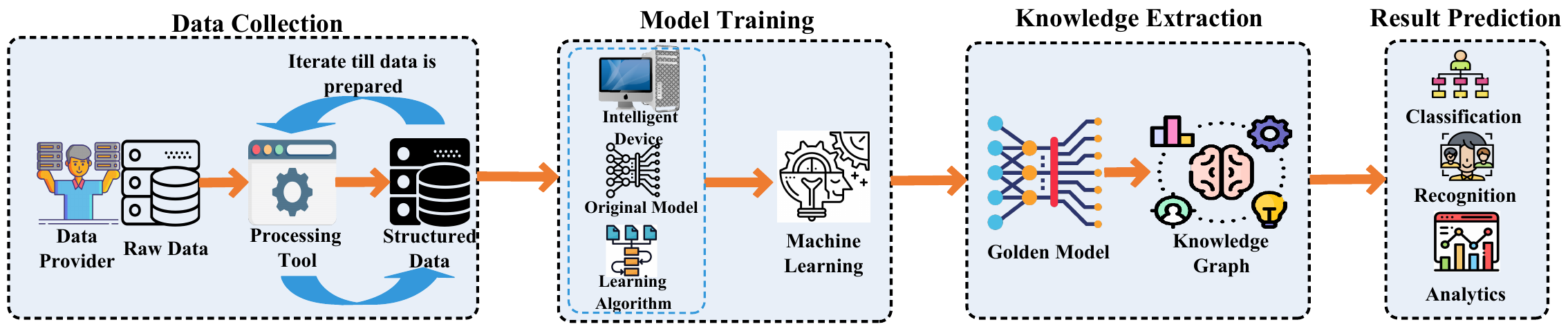}
    \caption{The process of machine learning in four key steps: data collection, model training, knowledge extraction, and result prediction.}
    \label{fig:ML}
\end{figure*}
\subsection{Background of Machine Learning}

Generally speaking, the core idea of ML algorithms can be summarized as training the machine to learn rules or patterns underlying some phenomenon using data and then making decisions or inferences based on new data using the learned rules or patterns. Many ML algorithms fall into the category of pattern recognition (PR), including face recognition, voice recognition, character recognition, and so on\cite{Schalkoff2007Pattern}. Since humans cannot easily program machines to follow all detailed rules and judgments, ML can be used to help machines learn hidden and even implied rules by themselves. This process is described simply as follows.

Suppose we are going to train a machine to classify whether a fruit is an apple or a banana (a classification task). We first collect some samples that can be labeled and learned by the machine (dataset). So some apples and bananas from this dataset along with their features, including shape, color, weight, size, and so on are recorded.  Now, a labeled fruit (apple or banana) with a set of ground-truth features together builds up a sample, and these labeled samples constitute the training dataset. The goal of this ML task is to make the machine learn features from the training dataset and output good predictions given new samples without labels (test dataset). This learning process can be expressed as fitting a function that takes the features as inputs and outputs a value that is as close as possible to the true label. Fig.~\ref{fig:ML} illustrates the procedure of ML with four main steps listed as follows:
\begin{itemize}
  \item \textbf{Data collection}. The quantity and quality of the collected data dictate how accurate the model is, and the dataset can be divided into training, validation, and test dataset \cite{wiki2021}.
  \item \textbf{Model training}. For different ML tasks, an appropriate model should be chosen wisely first. Then, the training dataset with the right labels is fed as inputs to the model to start training.
  \item \textbf{Knowledge extraction}. During training, features of the input samples are extracted by some metrics or combinations of metrics (e.g. linear or nonlinear combinations), and this knowledge helps the model updates its weights in structures.
  \item \textbf{Result prediction}. The test dataset which has been withheld from the model is used and outputs the prediction results, such as labels, values, vectors (e.g., generative time series), and matrices (e.g., generative images).
\end{itemize}

\subsection{Background of Distributed Machine Learning}
\begin{figure*}
    \centering
    \includegraphics[width=0.95\textwidth]{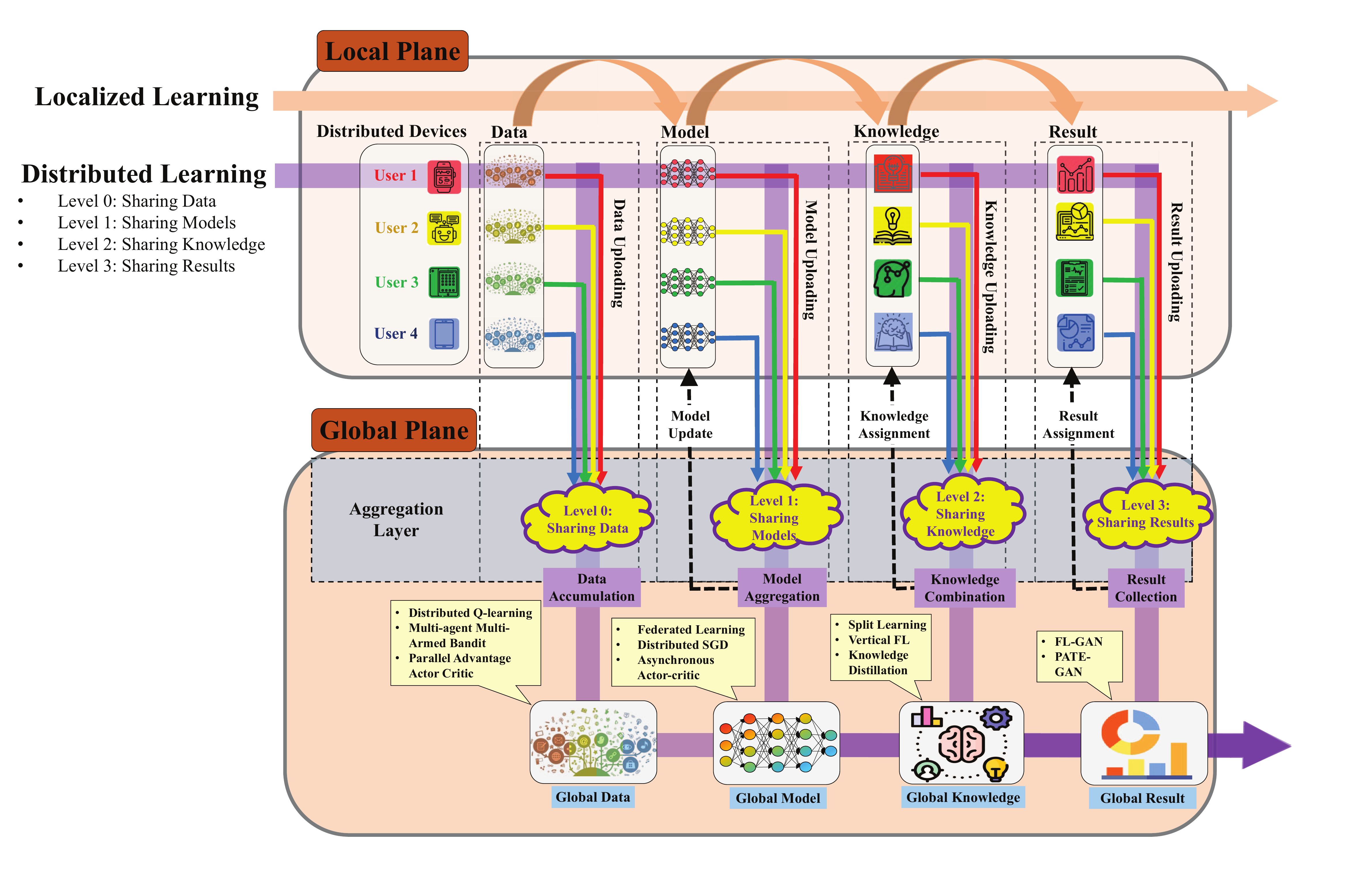}
    \caption{The framework of distributed learning, which is composed of a local and global plane. In the local plane, different information, i.e., data or models, are processed and generated in local devices, and then transmitted to a centralized server for aggregation. Four levels of the proposed distributed learning framework are described in detail, i.e., sharing data, sharing models, sharing knowledge, and sharing results, which are exemplified by representative ML techniques.}
    \label{fig:DML}
\end{figure*}

Distributed ML systems and algorithms have been extensively studied in recent years to scale up ML in the presence of big data. Existing work focuses either on the theoretical convergence speed of proposed algorithms or on the practical system aspects to reduce the overall model training time \cite{xing2016strategies}. Bulk synchronous parallel algorithms (BSP) \cite{Zinkevich2010Parallelized} are among the first distributed ML algorithms. Due to the hash constraints on the computation and communication procedures, these schemes share a convergence speed that is similar to traditional synchronous and centralized gradient-like algorithms.  The Stale synchronous parallel (SSP) algorithm \cite{Ho2013More} is a more practical alternative that abandons strict iteration barriers, and allows the workers to be off synchrony up to a certain bounded delay. The convergence results have been developed for both gradient descent and stochastic gradient descent (SGD) \cite{Ho2013More,Lian2015Asynchronous,Benjamin2011Hogwild} as well as proximal gradient methods \cite{Mu2014Communication} under different assumptions of loss functions. In fact, SSP has become central to various types of currently distributed parameter server architectures \cite{Abadi2016TensorFlow,Trishul2014Project,Kevin2017Gaia,Mu2016DiFacto}. Depending on how the workload is partitioned \cite{xing2016strategies}, distributed ML systems can be categorized into four levels:
\begin{itemize}
  \item \textbf{Level 0: sharing data}. After collecting and pre-processing data locally, each UE will upload its private/anonymized data to a central server, and then the server will use this aggregated data to complete the learning task.
  \item \textbf{Level 1: sharing model}. Different from uploading data directly, each UE can train a local ML model using its own data and {\color{blue}share} the trained model with the server. Then the server will aggregate the collected model and re-transmit the global model to UEs for the next round of learning.
  \item \textbf{Level 2: sharing knowledge}. Different from sharing ML models, the extracted knowledge from training local data, such as the relationship between different attributes, is further shared.
  \item \textbf{Level 3: sharing result}. The task training is completely processed locally, and each UE only shares ML results/outputs to the central server.
\end{itemize}
The detailed framework of the four-level distributed ML is illustrated in Fig.~\ref{fig:DML}, which is composed of a local and global plane. In the local plane, different information, i.e., data or models, are processed and generated in local devices, and then transmitted to a centralized server for aggregation. Four levels of the proposed distributed learning framework are described in detail, i.e., sharing data, sharing models, sharing knowledge, and sharing results, which are exemplified by representative ML techniques.

\subsection{Existing Distributed Learning Frameworks}
In this subsection, we will introduce some popular distributed learning models in the literature, which includes federated
learning, split learning, SGD-based collaborative learning, and multi-agent reinforcement learning.

\subsubsection{Federated Learning}
\begin{figure}
    \centering
    \includegraphics[width=0.45\textwidth]{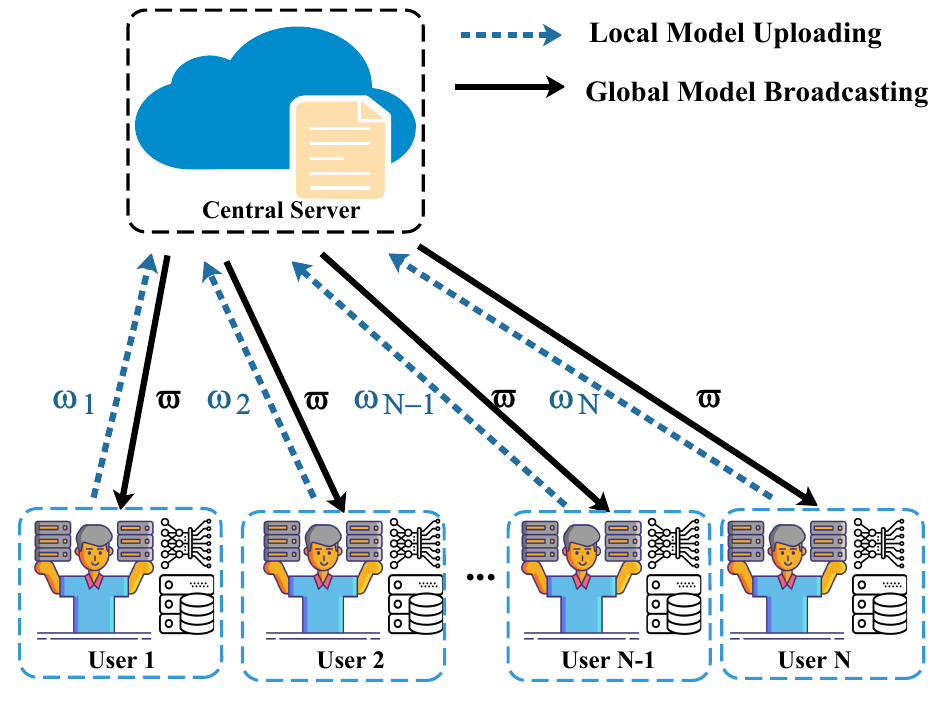}
    \caption{The structure of federated learning, where users train {\color{blue}{a}} ML model using their local data and share the models to a centralized server.}
    \label{fig:FL}
\end{figure}
FL is a collaborative ML technique \cite{mcmahan2017communication,konevcny2015federated,bonawitz2019towards} developed by Google,
which allows decoupling of data provision at UEs, and machine learning model aggregation,
such as network parameters of deep learning, at a centralized server. A structure of FL is plotted in Fig.~\ref{fig:FL}. The purpose of FL is to cooperatively learn a global model without directly sharing data. In particular, FL has distinct privacy advantages compared to data center training on a dataset. At a server, holding even an anonymized dataset can put client privacy at risk via linkage to other datasets. In contrast, the information transmitted for FL consists of minimal updates to improve a particular ML model. The updates can be ephemeral, and will not contain more information than the raw training data (by the data processing inequality). Further, the source of the updates is not needed by the aggregation algorithm, and so updates can be transmitted without identifying metadata over a mixed network such as Tor \cite{Chaum1981Untraceable} or via a trusted third party. General categories are distributed horizontal FL, where clients have different sample spaces with the same feature space, and share models during aggregation, distributed vertical FL with the same sample space and different feature spaces, sharing models or knowledge to the central server, and distributed transfer learning with various sample and feature spaces when uploading model or knowledge in aggregation \cite{Lim2020Federated}.

However, although the data is not explicitly shared in the original format, it is still possible for adversaries to reconstruct the raw data approximately, especially when the architecture and parameters are not completely protected. In addition, FL can expose intermediate results such as parameter updates from an optimization algorithm like SGD, and the transmission of these gradients may actually leak private information when exposed together with a data structure such as image pixels. Furthermore, the well-designed attacks such as inference attack (stealing membership information) \cite{melis2019exploiting,fredrikson2015model,hitaj2017deep}, and poisoning attack (polluting the quality of datasets or parameter models) \cite{bagdasaryan2020backdoor} may induce further security issues.
\subsubsection{Split Learning}
\begin{figure}
\centering
    \includegraphics[width=0.45\textwidth]{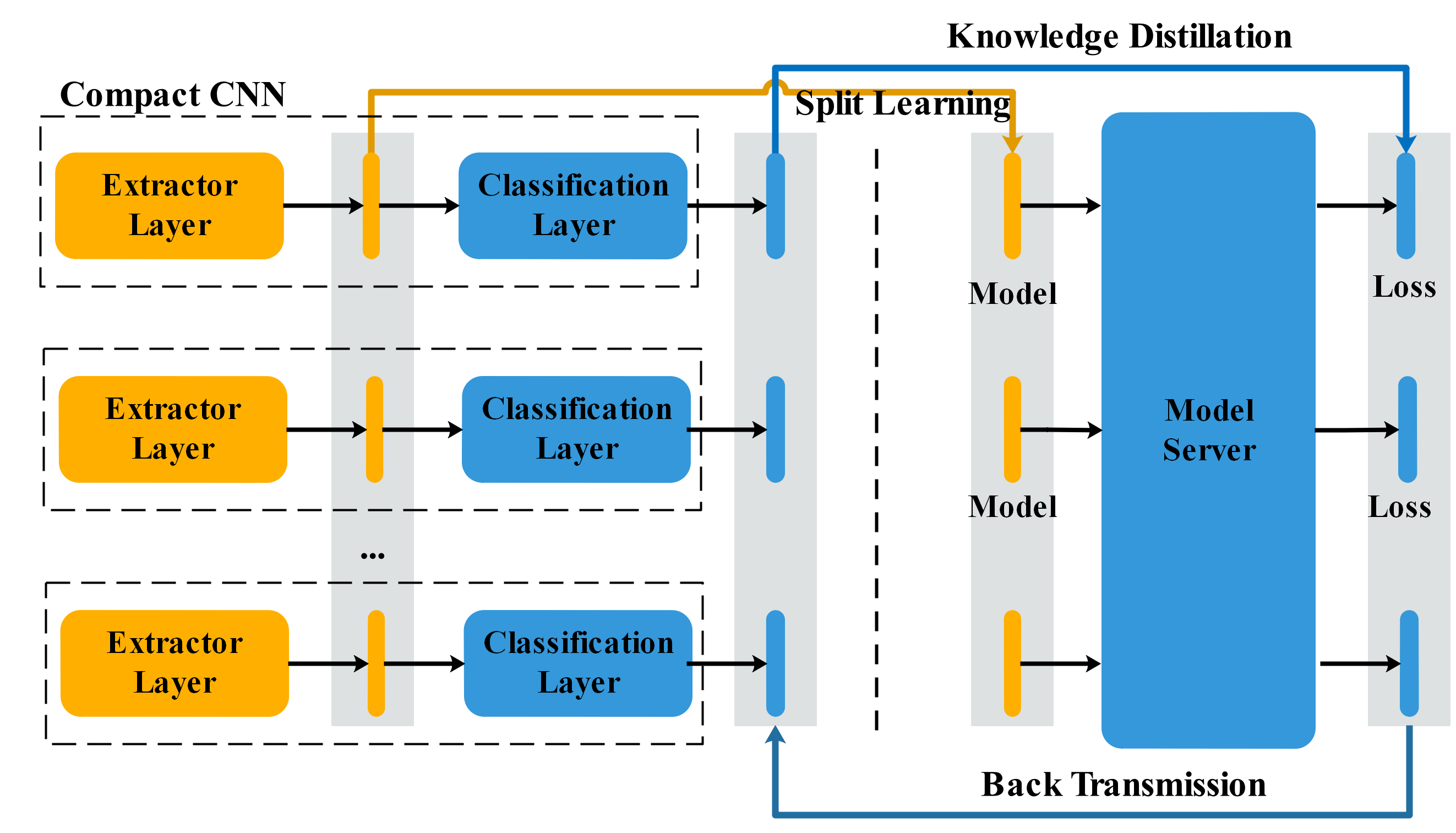}
    \caption{A reformulation of FL with assisted by the split learning and knowledge distillation \cite{He2020Group}.}
    \label{fig:split learning}
\end{figure}
Split learning, as a type of distributed deep learning \cite{vepakomma2018split,vepakomma2018no,vepakomma2019reducing,gupta2018distributed}, has another name of split neural network (SplitNN). Similar to FL, split learning is effective when data uploading is not available because of privacy and legal restrictions. In the SplitNN, each participant first trains a NN until a predefined layer, called the cut layer, and then transmits the output of the cut layer to the server. Upon receiving the outputs, a central server will continue training the rest layers. Then, the loss function value is calculated and back-propagated to the participant. When receiving the feedback, the participant continues the back-propagation until the network finishes training. In Fig.~\ref{fig:split learning}, we show a combination of FL and split learning, where the logits are shared and aggregated at a centralized server.

The computational and communication costs on the client side are reduced in split learning because part of the network is processed locally. In addition, instead of transmitting the raw data, the activation function of the cut layer is uploaded to the server, which has a relatively smaller size. Some experimental results show that split learning has higher performances and fewer costs than FL over figure classification tasks, i.e., CIFAR-100 datasets, using Resnet-50 architectures for hundreds of clients-based setups \cite{vepakomma2018split}. However, it needs further explanations on how split learning works and makes decisions, which is linked to the trust of distributed networks, especially in the health area \cite{vepakomma2018split}.

\subsubsection{Large Batch Synchronous SGD (LBS-SGD)}
The difference between the large batch synchronous SGD-based collaborative learning and FL lies in that the updates in LBS-SGD are processed on each batch of training data, and multiple epochs of local training are required before uploading in FL. In LBS-SGD, model parallelism and data parallelism are two common ways to support updating, such as distributed large mini-batch SGD \cite{goyal2017accurate}, distributed synchronous SGD with backups \cite{chen2016revisiting,vepakomma2018no}, and selective SGD \cite{shokri2015privacy}. In \cite{shokri2015privacy}, each participant is asked to choose a part of the models to update at each epoch and share them asynchronously with others. The work \cite{goyal2017accurate} considered synchronous SGDs by dividing local epochs into mini-batches over multiple clients and model aggregations. While the aggregated updates were performed synchronously in \cite{goyal2017accurate} that the aggregator will wait for all clients, the straggler may slow down the learning, and a synchronous optimization with backup participants has been provided in \cite{chen2016revisiting}.

\subsubsection{Multi-Agent Reinforcement Learning}
\begin{figure}
\centering
    \includegraphics[width=0.4\textwidth]{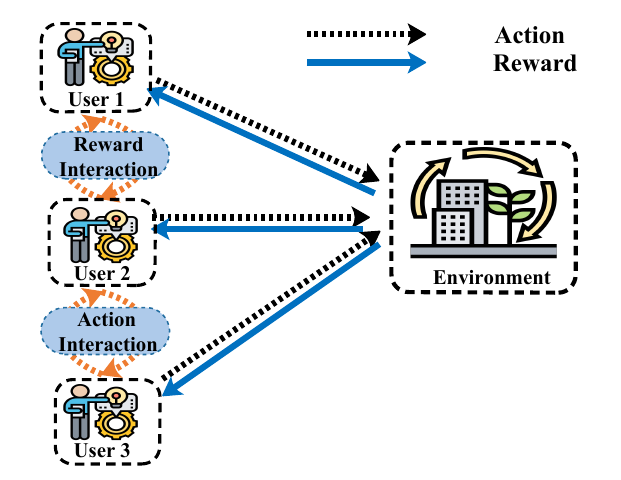}
    \caption{A framework of multi-agent reinforcement learning, where multiple users communicate and interact to change information, and also process actions to obtain feedback from the environment.}
    \label{fig:mrl}
\end{figure}
Reinforcement learning (RL) is trial-and-error learning by interacting directly with the environment, training according to the feedback, and finally achieving the designed goal. Specifically, RL defines a decision maker as an agent and the interaction with the environment, where three essential elements: the state, action, and reward, are used to describe the interaction. For each interaction, the client arrives at a certain state and processes a corresponding action, and then obtains feedback that is used to alter the current state to the next state. However, a single RL framework has no capability to address complex real-world problems, and thus, a multi-agent reinforcement learning system (MARL) has attracted increasing attention. Within a MARL, agents will cooperate with each other and observe the complex environment in a more comprehensive way. For example, as shown in Fig.~\ref{fig:mrl}, a three-agent reinforcement learning system, where actions and rewards are shared between different users, is provided. By absorbing the learning experiences from the user-self and other participants, a faster convergence rate with better performance is always achieved. However, compared to the single-agent setting, controlling multiple agents poses several additional challenges, such as the heterogeneity of participants, the design of achieved goals, and the more serious malicious client problem. Although a number of methods have been proposed to address these challenges, e.g., approximate actor-critic~\cite{Li2020F2A2} and lenient-DQN, limitations like nonseasonal communication among agents and privacy leakage prevent the rapid development of MARL and existing methods cannot be extended to large-scale multi-agent scenarios.

\begin{figure*}
\centering
    \includegraphics[width=0.9\textwidth]{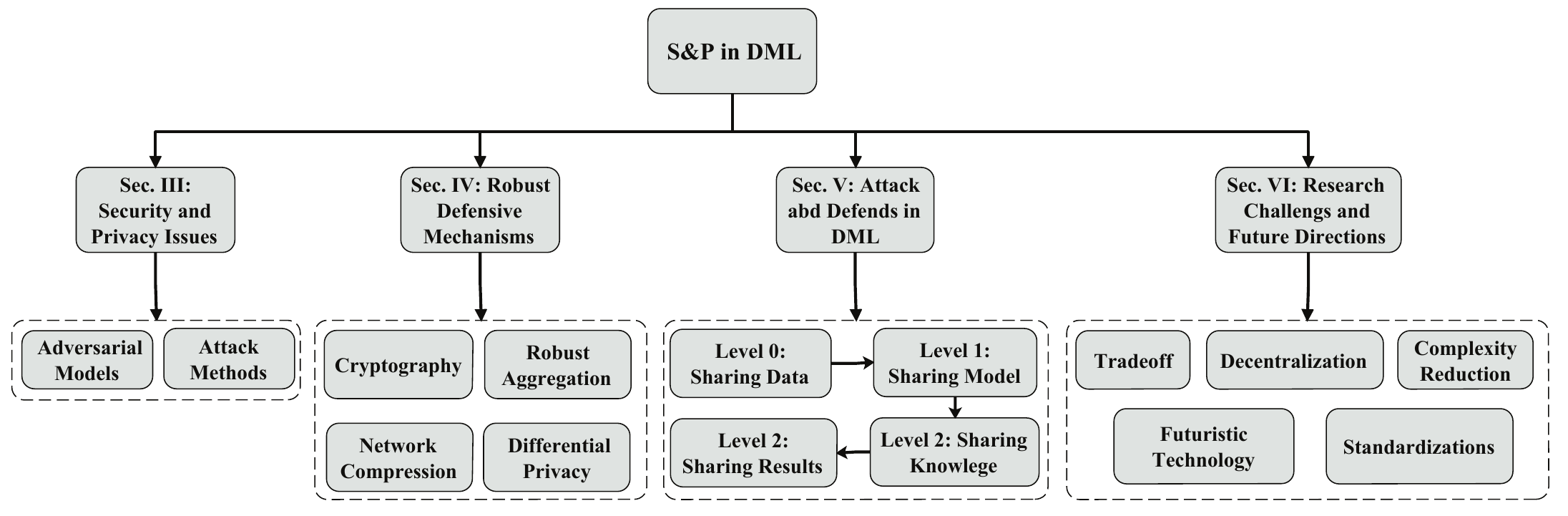}
    \caption{The structure of the survey with key compositions.}
    \label{structure}
\end{figure*}

Following the discussed background of distributed ML, we present the structure of this survey work in Fig.~\ref{structure}.
The rest of the paper is structured as follows. In Section~\uppercase\expandafter{\romannumeral3}, privacy and security issues are discussed and several robust protection methods are provided in Section~\uppercase\expandafter{\romannumeral4}. Then, in Section~\uppercase\expandafter{\romannumeral5}, we survey the attacks and  {\color{blue}defenses} in various paradigms in distributed ML. Several research challenges and future directions are shown in Section~\uppercase\expandafter{\romannumeral6}. Finally, conclusions are drawn in Section~\uppercase\expandafter{\romannumeral7}. In addition, a list of important abbreviations is provided in Table~\ref{ListofAbbr}.

\begin{table*}[hbt]
\caption{List of Important Abbreviations.}
\centering
\begin{tabular}{|l|l|l|l|l|l|}
\hline
\textbf{Abbr.}&\textbf{Definition}&\textbf{Abbr.}&\textbf{Definition}&\textbf{Abbr.}&\textbf{Definition}\\
\hline
ML& Machine Learning& DL & Deep Learning & RL & Reinforcement Learning\\
DQN & Deep Q-Learning & AC & Actor-Critic &A3C & Asynchronous Advantage Actor-Critic\\
TRPO& Trust Region Policy Optimization &PG &Policy Gradient &PPO&Proximal Policy Optimization\\
DP& Differential Privacy & HE & Homomorphic Encryption & SMC & Secure Multiparty Computation\\
SGD & Stochastic Gradient Descent & FL &Federated Learning & NN & Neural Network\\
\hline
\end{tabular}
\label{ListofAbbr}
\end{table*}
\section{Privacy and Security Risks in Distributed ML}
In Section~\uppercase\expandafter{\romannumeral3}, we will introduce the potential risks of privacy and security, which are measured by factors including threat models, adversarial models, and attack methods.
\subsection{Threat Models}
\subsubsection{Malicious/Curious Participant}
Participants in distributed ML can be malicious or curious. For example, a car insurance company with limited user attributes might want to improve its risk evaluation model by incorporating more attributes from other businesses, e.g., a bank, a taxation office, etc. The role of the other participants is simply to provide additional feature information without directly disclosing their data to other participants, and in return, obtain financial and/or reputation rewards. However, the competitors may be disguised as collaborators, and then damage the training process or steal the ML model.
\subsubsection{External Attackers}
In terms of exchanged information, eavesdropping, modification or deletion can occur during communication in distributed ML as well. We can notice that the exchanged information usually contains the updated direction and extracted features from private data, and thus it is crucial to ensure its correctness, especially for the client-server framework. An external attacker may control the final output by modifying or deleting the exchanged information in the communication. In addition, via eavesdropping on the extracted features from private data, an external attacker can further infer sensitive information~\cite{Shokri2017Membership}.

\subsection{Adversarial Models}
In this subsection, we will discuss adversarial goals related to leaking information from the training data or destroying models during learning.

\subsubsection{Access}
\begin{itemize}
\item  White-Box: The adversary is assumed to acknowledge certain information about the training data or the learning model, e.g., model parameters, network structures, or part of/the whole training dataset.
\item  Black-Box: The adversary does not have any knowledge about the ML model, but the adversary can further explore the model by injecting some designed inputs and observing related outputs \cite{papernot2017practical}.
\end{itemize}

\subsubsection{Training v.s. inference}
The second factor is the place where the attack happens:
\begin{itemize}
\item Training Stage: The adversary attempts to learn the model by accessing a part or all of the training data, and creating a substitute model, i.e., a shadow model.
\item Inference Stage: The adversary observes the outputs from the learning and sums up the model characteristics~\cite{Shokri2017Membership}.
\end{itemize}

\subsubsection{Passive vs. Active}
A third factor is to distinguish between passive and active attacks.
\begin{itemize}
\item Passive attack: The adversary can passively observe and obtain the updates but change nothing during the training process.
\item Active attack: The adversary actively performs and adjusts the learning operation. For example, the adversary can upload unreasonable parameters to degrade the aggregate model in FL~\cite{9429701}.
\end{itemize}

\subsection{Attack Methods}
In this subsection, several attack methods are investigated as follows.
\subsubsection{Poisoning Attack}
The goal of a poisoning attack is to degrade the model quality, which misleads the learning to an incorrect direction by carefully crafting poisoning samples during training, also called adversarial examples \cite{goodfellow2015explaining}. In the black-box attack, the attacker can only inject a relatively small amount of crafted/poisoned data into the training model, where the amount and the undiscovered capability of these poisoning data are two basic metrics to estimate the attacking performance. For example, the authors in \cite{Jagielski2018Manipulating} have first investigated poisoning attacks against linear regression models and proposed a fast optimization algorithm with limited crafting samples to perturb outputs. Further, Suciu \emph{et al.} have investigated the minimum information required by the attacker to achieve various attacking goals \cite{Octavian2018When}. In the white-box attack, the adversaries have full knowledge of the training model and can take advantage of it to reconstruct a powerful poisoning attack. For example, Yuan \emph{et al.} in \cite{Yuan2019Adversarial} have proposed a white-box attack with perfect knowledge under different goals. Although the mentioned method might be unrealistic in practical settings, it can achieve almost five times than the black-box attack in success rate.
\subsubsection{Evasion Attack}
An evasion attack often happens in the prediction process, which aims to mislead the outputs. In detail, the evasion attack is to change real data from one category to a determined or random one and destroy the integrity of the original dataset. From a black-box attack angle, the adversary only knows the type of the training dataset and observes the outputs. Based on this assumption, the authors in \cite{Kwon2020Selective} have realized it  {\color{blue}in} the speech recognition system. The generated adversarial samples achieve a $91.67\%$ successful rate on moving one data from one category to another. While in the white-box attack, the adversary is able to acknowledge more useful information, such as the network structure and the type of training samples, rather than the predictive interface. For example, Kevin Eykholt \emph{et al.} in \cite{Eykholt2018Robust} has shown the weakness for DNNs when random noises are added to the inputs and an advanced robust physical perturbations-based method has been proposed.
\subsubsection{Model Inversion Attack}
The model inversion attack proposed in \cite{fredrikson2014privacy} works in a black-box fashion, and the adversary only knows the input and can observe the corresponding outputs, which is used to detect correlations between uncertain inputs and respective outputs. A follow-up work has presented a combination with a black-and-white box attack \cite{fredrikson2015model}. The proposed attack aims to predict the highest probability of one input for a given label, which the adversary is able to reconstruct the input for a known label, i.e., a figure from a specific class. However, the proposed model inversion attack only works in linear models for most cases, and a major weakness is that the complexity grows exponentially with the input size since it relies on searching all linear combinations by brute force.
\subsubsection{Membership Inference Attack}
The membership inference attack (MIA) is mainly focused  {\color{blue}on privacy attacks}. A previous attack targeting distributed recommend systems \cite{calandrino2011you} intended to infer which input will lead to a change in the output by observing temporal patterns from the learning model. In \cite{Shokri2017Membership}, Shokri \emph{et al.} have investigated the differences between the models to infer whether an input exists in the training dataset for the supervised model. In particular, a shadow model analogs as a similar structure to the targeted model in a black-box fashion. Following \cite{Shokri2017Membership}, Song \emph{et al.} in \cite{song2017machine} attempted to record the training data with black-box access. Then, the authors in \cite{ateniese2015hacking} have exploited the knowledge of learning models to hide the Markov model and attack support vector machine in classification tasks. Also, related works \cite{aono2017privacy,hitaj2017deep,melis2018inference} presented inference attacks against distributed deep learning \cite{mcmahan2017communication,shokri2015privacy}. In particular, Aono \emph{et al.} \cite{aono2017privacy} aimed to attack the privacy-preserving learning framework proposed in \cite{shokri2015privacy}, and revealed that partial data samples can be revealed by an honest-but-curious server. However, the operation that the single-point batch size limits its effectiveness. Also, a white-box attack against \cite{shokri2015privacy} has been proposed in \cite{hitaj2017deep}, which used generative adversarial networks (GAN) to produce similar samples with a targeted training dataset, however, the proposed algorithm lost effectiveness in the black-box access. Finally, Truex \emph{et al.} in \cite{truex2018towards} has shown that the MIA is usually data-driven, and Melis \emph{et al.} in \cite{melis2018inference} have demonstrated the way that a malicious participant infers sensitive properties in distributed learning. Other MIAs focused on genomic research studies \cite{backes2016membership,homer2008resolving}, in which the attack is designed to infer the presence of specific information of individuals within an aggregated genomic dataset \cite{homer2008resolving}, locations \cite{pyrgelis2017knock}, and noisy statistics in general \cite{dwork2015robust}.

\subsubsection{Model and Functionality Stealing}
\begin{itemize}
\item Model Extraction.
The aim of model extraction is first proposed in \cite{tramer2016stealing}, in which they proposed to infer the parameters from a trained classifier with a black-box fashion; however, it only works when the adversary has access to the predictions, i.e., the probabilities for each class in a classification task. In follow-up works, other researchers went a step further to perform hyper-parameter stealing \cite{wang2018stealing}, which are external configurations. These values cannot be estimated by data samples, architecture extraction \cite{oh2019towards} that infers the deep model structures as well as the updating tools (e.g., SGD or alternating direction method of multipliers (ADMM)), etc.
\item Functionality Extraction.
The concept of functionality extraction is, rather than stealing the model, to create knock-off models. Orekondy \emph{et al.} \cite{orekondy2019knockoff} have processed this attack only based on design inputs and relative outputs to observe correlation from ML as a service (MLaaS) queries. In particular, the adversary uses the input-output pairs, e.g., image-prediction pairs in a figure classification task, to train a knock-off model, and compares it with one of the victims for the same task. In addition, the authors in \cite{papernot2017practical} have trained a shadow model to replace a DNN which directly uses inputs generated by the attacker and labeled by the attacking DNN.
 \end{itemize}

\subsection{Section Summary}
To sum up, the attack target can be regarded as a clue to distinguish the privacy and security risks from the adversary aspect. A common aim for the privacy attack is to infer a membership of participants without degrading the learning performance, i.e., membership inference attack, and model and functionality stealing, while malicious clients usually aim to destroy the integrity of the learning system, i.e., model poisoning, evasion, and inversion attack.
\section{Robust Defensive Mechanisms}

In Section~\uppercase\expandafter{\romannumeral4}, we will present an overview of several robust defensive mechanisms that include cryptography, robust aggregation, network compression, and differential privacy to reduce information leakage and address security issues.
\subsection{Cryptography}
Cryptography is a vital part of distributed ML as it has the ability to support confidential secure computing scenarios. There are a vast of research algorithms and prototypes in literature, which allow participants to obtain learning outputs without uploading their raw data to the server. For instance, in the supervised ML task, secure multi-party computation (SMPC) and homomorphic encryption (HE) based privacy-enhancing tools have been proposed to enable secure computing. Typical examples are, neural networks \cite{bonawitz2017practical,Liu2017Oblivious,mohassel2017secureml}, matrix factorization \cite{Nikolaenko2013Privacy}, linear regressions \cite{Du2004Privacy}, decision trees \cite{bost2015machine}, and linear classifiers \cite{Lauter2014Private,Graepel2012ML}.

Specifically, SMPC allows two or more participants to jointly complete a ML task over the shared data without revealing it to others. Popular SMC prototypes are usually developed for two parties, such as  \cite{Du2004Privacy,mohassel2017secureml,kilbertus2018blind,li2018privpy} designed for distributed ML tasks. For more than two parties, algorithms based on three-party communication have been provided in \cite{mohassel2018aby3,araki2016high,mohassel2015fast,furukawa2017high}, which all rely on the majority of semi-honest or honest participants. For example, Bonawitz \emph{et al.} in \cite{bonawitz2017practical} has proposed a mixture of several communicating schemes to enable secure computing of participants in FL by blurring the aggregation from the server.

Regard to HE, it mainly uses the encryption and decryption protocol to transform the original message by certain mathematical operations, and there are three common forms for HE: 1) Partially Homomorphic Encryption (PHE) supports one type of mathematical operation; 2) Somewhat Homomorphic Encryption (SWHE) that uses a number of mathematical operations for limited use-cases; 3) Fully Homomorphic Encryption (FHE) supports unlimited numbers of mathematical operations with no other limits \cite{acar2018survey}. For example, Phong \emph{et al.} in \cite{aono2017privacy} have developed a novel homomorphic scheme based on additive operations for FL with no performance degradation \cite{aono2017privacy}. Other distributed learning strategies, such as \cite{zhang2015privacy,yuan2013privacy} used HE to encrypt data and the central server can train a learning model based on the encrypted one.
However, the drawbacks of HE are obvious. First, it is usually hard or even impractical to implement HE since this will generate a huge computation overhead \cite{li2018privpy,Abbas2018A,Lu2017Using}.
Second, with the increasing number of homomorphic operations, the size of the encrypted models grows exponentially, especially in the SWHE \cite{Abbas2018A}, which usually largely surpasses the original model. Third, extra communications between the client and server are required to facilitate key-sharing protocols, which will increase communication costs.
\subsection{Robust Aggregation}
The robust aggregation protection methods are used designed for distributed ML that a server needs to aggregate something from clients. To prevent malicious clients, or a group of collusive malicious clients, such as the Byzantine attack in FL \cite{Lamport2019The}, the authors in \cite{Blanchard2017Machine} have proposed Krum, a robust aggregation scheme. By minimizing the sum of squared Euclidean distances over the aggregated models, Krum can effectively recognize and remove these outliers. Several follow-ups~\cite{Mhamdi2018The,Li2014Resolving,Pillutla2019Robust} aimed to recognize malicious clients. In addition, Chang \emph{et al.} \cite{Chang2019Cronus} have developed a knowledge-sharing-based algorithm to preserve privacy. The proposed Cronus algorithm relies on a public dataset that is available to all clients. Instead of transmitting parameters, clients will upload the predicted results from this public dataset, and a mean estimation algorithm \cite{Diakonikolas2017Being} was used to aggregate these high dimensional label samples. Although Cronus has been proven to defend against basic model poisoning attacks with an acceptable performance loss, sharing labels will lead to privacy leakage to a certain extent.
\subsection{Network Compression}
The main purpose of compressing the network is to reduce information transmission, which saves communication resources and accelerates learning. As well, it can also reduce the information exposed to the adversary. Typical methods include quantization \cite{Alistarh2017QSGD,Bernstein2018signSGD,Wei2017TernGrad}, network sparsification \cite{Alistarh2018The,Sebastian2018Sparsified}, knowledge distillation \cite{lin2013network,jeong2018communication}, network pruning \cite{Liu2019Rethinking,Jiang2022Model} and Sketch \cite{haddadpour2020fedsketch,Jiang2018SketchML,Li2019Privacy}. Specifically,
an initial work \cite{shokri2015privacy} provided the ideal to transmit a subset of all gradients in distributed SGD, and based on it, the authors in \cite{yoon2020federated} have proposed a novel gradient subset scheme that uploads sparse and chosen gradients can improve the prediction accuracy in the non-independent and identically distributed (non-IID) settings. However, as the gradients keep their own form, recent works \cite{Phong2018Privacy,melis2019exploiting} shown that such methods cannot prevent a specific adversary from inferring available information from these frameworks \cite{Phong2018Privacy,melis2019exploiting}.

Another approach is using lossy compression techniques to decrease the transmitted bits, and it may facilitate certain forms of information security. The authors in \cite{Reisizadeh2020FedPAQ} quantized the updates using the low-precision quantizer proposed in \cite{Alistarh2017QSGD} and provided a smooth tradeoff between compression rate and the convergence performance in convex and non-convex settings. In \cite{Rothchild2019FetchSGD}, a count Sketch method with momentum and error accumulation was provided for FL while achieving a high compression rate with good convergence. On the basis of it, the authors in \cite{Li2019Privacy} have proved such a quantization method can provide a certain differential privacy guarantee. Moreover, a Sketch-based method was proposed in \cite{Jiang2018SketchML}, which sorts gradient values into buckets and encodes them with bucket indexes. In addition, a stochastic-sign-based gradient compressor was used and analyzed to enable communication efficiency \cite{Jin2020Stochastic}, and an auto-encoder compressor was proposed in \cite{li2019end} in which the auto-encoder is trained based on dummy-gradients, and the server will release the coded part to clients while keeping the decoder part secretive.

Different from the above methods, a technique called dropout can also be used to defend \cite{hinton2012improving}, although it is usually used to prevent overfitting problems in training \cite{srivastava2014dropout}. By applying dropout, there will be no deterministic outputs (e.g., the updating gradients) on the same training dataset, which can reduce the exploitable attack fact \cite{melis2019exploiting}.
\subsection{Differential Privacy}
Differential privacy (DP) is a standard definition for privacy estimation \cite{dwork2008differential}.   {\color{blue}A} query mechanism is first defined as a property to a dataset, DP-based analytical methods are then extended for ML models on private training data, such as SVM \cite{rubinstein2009learning}, linear regression\cite{zhang2012functional}, and deep learning \cite{shokri2015privacy,abadi2016deep}. On neural networks, differentially private stochastic gradient descent \cite{abadi2016deep} is the most famous method that adds random noises on the updating gradients to achieve DP guarantee.

DP sets up a game where the adversary is trying to determine whether a training model has an input $D$ or $D'$ which are adjacent datasets and only differ in one sample. If the adversary can distinguish which dataset ($D$ or $D'$) is used to train by observing the outputs, we can say this training model leaks private information. A formal definition of ($\epsilon, \delta$)-DP is expressed as follows:
\begin{defn} \label{dp}
$\left((\epsilon, \delta)-\textrm{DP}\right)$. A randomized mechanism $f: \textrm{\textbf{D}} \mapsto \mathcal{R}$ offers ($\epsilon, \delta$)-DP if for any adjacent input $d, d' \in \textrm{\textbf{D}}$ and $S \subset \mathcal{R}$,
\begin{equation}
\Pr \left[ {f\left( d \right) \in S} \right] \le {e^\epsilon }\Pr \left[ {f\left( {d'} \right) \in S} \right] + \delta,
\end{equation}
where $f(d)$ denotes a random function of $d$.
\end{defn}
To estimate the accumulated privacy budget in multiple learning iterations, the composition theory in \cite{dwork2008differential} shown the effectiveness, and other variants of DP \cite{Mironov2017Renyi,Dwork2016Concentrated} use slightly different formulations with (\ref{dp}), and can achieve a tighter privacy delimitation. Recently, the authors in \cite{Milad2021Adversary} have derived a lower bound of DP from the adversary perspective, and the Monte Carlo-based method is the first trial of obtaining the privacy level empirically. In addition, the concept of local DP was proposed firstly in \cite{duchi2013local,Wang2017Locally}, and enjoys its popularity gradually.

\subsection{Section Summary}
To sum up, general defensive schemes, such as cryptography, robust aggregation, and network compression, can provide thorough protection on security and preserve privacy, where the application of DP is particularly for privacy issues.

\section{Attacks and Defences in Various Levels of Distributed Learning}
In Section~\uppercase\expandafter{\romannumeral5}, we will provide a detailed discussion on the state-of-the-art of attacks and defenses in each level of distributed ML.
\subsection{Level 0: Sharing Data}
Data collection plays an important role in various data-governed distributed ML algorithms.
However, original data usually  {\color{blue}contain} sensitive information such as medical records, salaries, and locations, and thus a straightforward release of data is not appropriate.
Correspondingly, research on protecting the privacy of individuals and the confidentiality of data with an acceptable performance loss has received increasing attention from many fields such as computer science, statistics, economics, and social science.
\begin{figure*}
\centering
    \includegraphics[width=0.95\textwidth]{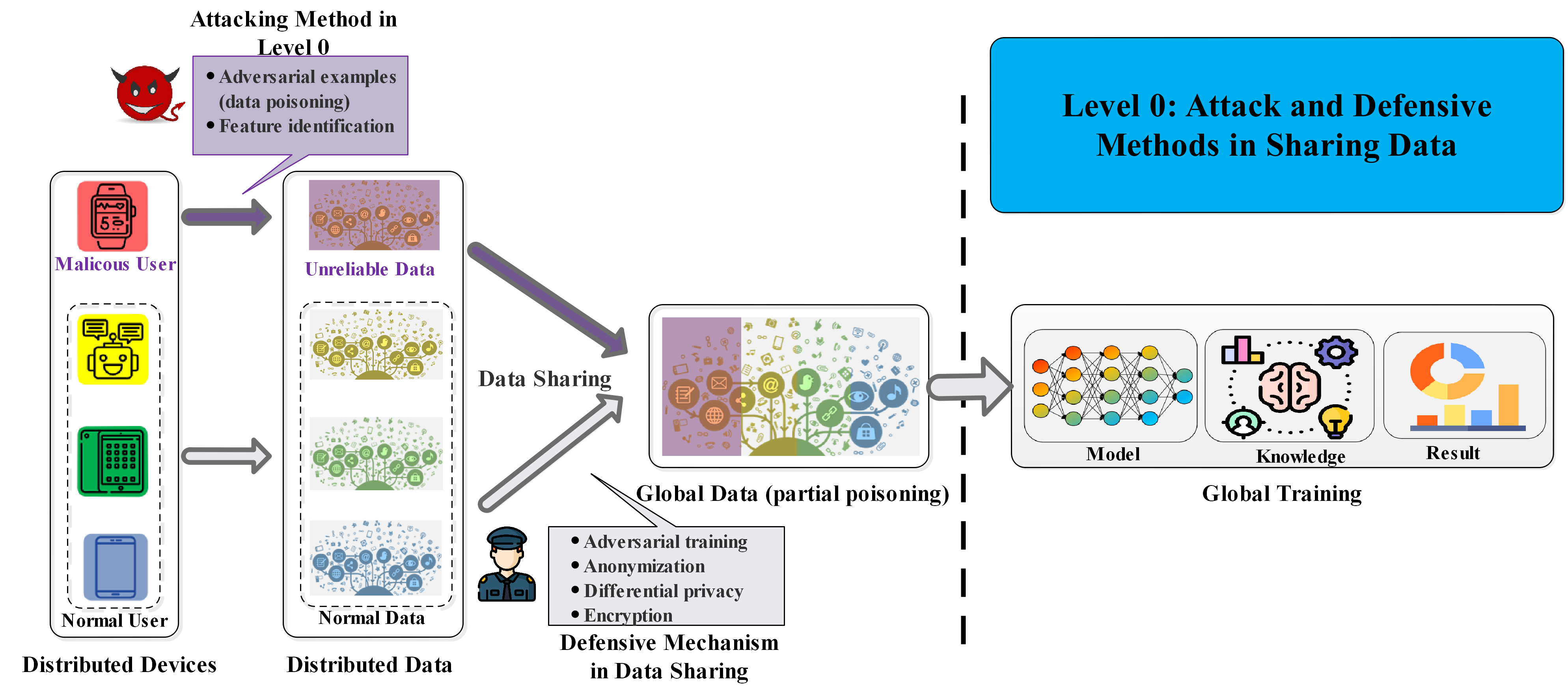}
    \caption{A breakout figure from Fig.~\ref{fig:DML}: an illustration of privacy and security issues in Level 0 distributed learning with sharing data.}
    \label{fig:data-agg}
\end{figure*}
\subsubsection{Threat Models}
Although existing works have proposed a mount of mechanisms to hide identifiers of the raw data, it is also possible for attackers to steal privacy by analyzing hidden features~\cite{Raymond2010Privacy}.
Moreover, deep neural networks have been proven vulnerable to adversarial examples, which poses security concerns due to the potentially severe consequences~\cite{Dong2018Boosting}.
This means that if some adversaries successfully make adversarial examples participate in system training, the training performance will be unacceptable.
\subsubsection{Taxonomy of Attacks}
Attacks on data publishing models can be mainly categorized as adversarial examples and feature identification based on their goals. As shown in~Table~\ref{tab:attacks_data_sharing}, we summarize possible attacks as follows.
\begin{itemize}
\item[$\bullet$] \textbf{Adversarial examples (data poisoning).}
The work in~\cite{Dong2018Boosting} integrated the momentum term into the iterative process for attacks and generated more transferable adversarial examples by stabilizing update directions and escaping from poor local maxima during the generating iterations.
The research on this area is faced with an ``arms race'' between attacks and defenses, i.e., a defense method proposed to prevent the existing attacks will be soon evaded by new attacks.
\item[$\bullet$] \textbf{Feature identification.} Although many works have proposed efficient methods to process original data in order to preserve sensitive information.
Many feature identification attacks are emerging to expose hidden information.
As one of the feature identification attacks, structure-based de-anonymization attacks to graph data have been proposed, which aims to de-anonymize the private users in terms of their uniquely distinguishable structural characteristics~\cite{Ji2017Graph}.
\end{itemize}
\begin{table*}[hbt]
\caption{Taxonomy of attacks in Level-0 distributed ML with sharing data.}
\centering
\begin{tabular}{|m{1.5cm}<{\centering}||m{1.5cm}<{\centering}|m{4cm}<{\centering}|m{4cm}<{\centering}|m{4cm}<{\centering}|}
\hline
\textbf{Issue}& \textbf{Ref.}& \textbf{Attacker's knowledge}& \textbf{Learning Model}& \textbf{Effectiveness}\\
\hline\hline
\multirow{8}*{\shortstack{Adversarial\\
examples}}
&\cite{Dong2018Boosting}& White-box, black-box & Inception v2, Inception v3, Inception v4, Resnet v2-152& Attack a white-box model with a near $100\%$ success rate and more than $50\%$ for black-box models\\
\cline{2-5}
&\cite{Sandy2017Adversarial}&White-box, black-box&DQN, A3C, TRPO&Physically interfering with the observations of the victim\\
\cline{2-5}
&\cite{Xiao2019Characterizing}&Black-box&AC&Directly attack actions to achieve the designated purposes\\
\cline{2-5}
&\cite{Zhong2020Adversarial}&Black-box&AC&Taking actions to induce natural observations (environment dynamic) that are adversarial to the victim\\
\hline\hline
Feature
identification
&\cite{DArvind2008Robust}&A little bit about an individual subscriber&-&Identify the Netflix records of known users, uncovering users' preferences and other sensitive information\\
\hline
\end{tabular}
\label{tab:attacks_data_sharing}
\end{table*}
\subsubsection{Taxonomy of Defences} Many defensive mechanisms have been designed against aforementioned attacks as shown in Table~\ref{tab:defences_data_sharing}, and we will discuss various defenses as follows.
\begin{itemize}
\item[$\bullet$] \textbf{Adversarial training.}  Adversarial training is among the most effective techniques to improve model robustness by augmenting training data with adversarial examples.
The work in~\cite{Dong2020Adversarial} has proposed an adversarial distributional training (ADT) framework, which is formulated as a mini-max optimization problem and improves the model robustness obviously.
In this framework, the inner maximization aims to learn an adversarial distribution to characterize the potential adversarial examples around a natural one under an entropic regularizer, and the outer minimization aims to train robust models by minimizing the expected loss over the worst-case adversarial distributions.
\item[$\bullet$] \textbf{Anonymization.} An anonymization operation comes in several flavors: generalization, suppression, anatomization, permutation, and perturbation~\cite{Sweeney2002K,Shaham2020Privacy}.
These techniques aim to remove or hide identifying characteristics from raw data while guaranteeing the data utility.
An information-theoretic approach has been formulated and proposed a new multi-objective loss function for training deep auto-encoders~\cite{Mohammad2019Mobile}, which helps to minimize user-identity information as well as data distortion to preserve the application-specific utility.
The work in~\cite{Maximov2020CIAGAN} has proposed the conditional identity anonymization generative adversarial
networks (CIA-GAN) model, which can remove the identifying characteristics of faces and bodies while producing high-quality images and videos that can be used for various computer vision tasks, such as detection or tracking.
Unlike previous methods, CIA-GAN has full control over the de-identification (anonymization) procedure, ensuring both anonymization as well as diversity.
In summary, the choice of anonymization operations has an implication for the search space of anonymous tables and data distortion.
The full-domain generalization has the smallest search space with the largest distortion, and the local recording scheme has the largest search space but the least distortion.
\item[$\bullet$] \textbf{Dummy.}
Existing research methods to protect data privacy mainly focus on the  protection of the user's identities through anonymity. User attributes can be classified into identity information, quasi-identifier, and sensitive information. Given an anonymity table, if the attributes in the table have not been properly treated, an adversary may deduce the relationship between the user's identity and sensitive information according to the user's quasi-identifier, such as age and gender.
A popular approach for data anonymity is $k$-anonymity, and any record in a $k$-anonymized dataset has a maximum probability $1/k$ of being re-identified \cite{samarati1998protecting,Samarati2001Protecting,el2008protecting}.
The privacy model $l$-diversity and $t$-closeness in \cite{li2007t} further refines the concept of diversity and requires that the distribution of the sensitive values of each equivalent class should be as close as to the overall distribution of the dataset. The common rules for these algorithms are basically to produce dummy records to hide the real ones.
In addition, the dummy-based methods also work for location privacy protection. Dummy data along with the true one will be sent to the server from users, which may hide the client's contribution during training~\cite{Sina2020Privacy}. Because the collection is processed on the server, the system performance can still be guaranteed.
As an efficient method to generate realistic datasets, GANs provide an alternative to balance user privacy and training performance.
The work in~\cite{Shaker2020Generalization} has proposed a novel data augmentation technique based on the combination of real and synthetic heartbeats using GAN to improve the classification of electrocardiogram (ECG) heartbeats of 15 different classes from the MIT-BIH arrhythmia dataset\footnote{https://www.physionet.org/content/mitdb/1.0.0/}.
\item[$\bullet$] \textbf{DP.}
As a promising solution, a mechanism is said to be differentially private \cite{dwork2008differential} if the computation result of a dataset is robust to any change of the individual sample. Several differentially private machine learning algorithms \cite{ji2014differential} have been developed in the community, where a trusted data curator is introduced to gather data from individual owners and honestly runs the private algorithms. Compared to DP, Local DP (LDP) \cite{duchi2013local,Wang2017Locally} eliminates the need for a trusted data curator and is more suitable for distributed ML.
Rappor \cite{erlingsson2014rappor}, which applies LDP by Google, is designed to collect the perturbed data samples from multiple data owners. Besides simple counting, a follow-up paper \cite{fanti2016building} shows that Rappor can also compute other types of statistics such as joint-distribution estimation and association testing.
Besides Rappor, an alternative way that achieves DP is to add random noise on the sample value before publishing \cite{duchi2013local,dwork2006calibrating}. To process this method, a numerical sample is always normalized and a categorical one is transformed to the same range by one-hot coding.
In addition, the authors in~\cite{Balcan2012Distributed} adopted the DP algorithm to handle the privacy concern in a communication problem that each distributed client needs to transmit data to one aggregated center to learn a model.
The work\cite{Omobayode2020Efficient} has proposed a distributed edge computing which for image classification, where each edge will upload its raw data after coding to latent data to protect privacy.
\item[$\bullet$] \textbf{Encryption.}
The work in~\cite{David2021Scalable} has instantiated  a scalable privacy-preserving distributed learning (SPINDLE), an operational distributed system that supports the privacy-preserving training and evaluation of generalized linear models on distributed datasets.
Moreover, it relies on a multiparty HE scheme to execute high-depth computations on encrypted data without significant overhead.
The work in~\cite{Xu2015Privacy} has proposed a distributed algorithm for distributed data, where privacy is achieved by the data locality property of the Apache Hadoop architecture and only a limited number of cryptographic operations are required.
\item[$\bullet$] \textbf{Others.}
The work in~\cite{Zhang2015Secure} has aimed to develop secure, resilient, and distributed ML algorithms under adversarial environments.
This work has established a game-theoretic framework to capture the conflicting interests between the adversary and a set of distributed data processing units.
The Nash equilibrium of the game has allowed for predicting the outcome of learning algorithms in adversarial environments and enhancing the resilience of the ML through dynamic distributed learning algorithms.
\end{itemize}

\begin{table*}[hbt]
\caption{Taxonomy of defenses in Level-0 distributed ML with sharing data.}
\centering
\begin{tabular}{|m{2cm}<{\centering}||m{1.5cm}<{\centering}|m{3.5cm}<{\centering}|m{4cm}<{\centering}|m{4cm}<{\centering}|}
\hline
\textbf{Method}& \textbf{Ref.}& \textbf{Use case}& \textbf{Key idea}& \textbf{Effectiveness}\\
\hline\hline
\multirow{1}*{\shortstack{Adversarial\\
training}}
&\cite{Dong2020Adversarial}& Against adversarial examples &Formulating a minimax optimization problem, Parameterizing the adversarial distributions&Improving model security and robustness\\
\hline\hline
\multirow{5}*{\shortstack{Anonymization}}
&\cite{Shaham2020Privacy}&Removing unique identifiers of spatiotemporal trajectory datasets&Clustering the trajectories using a variation $k$-means algorithm&Enhancing the $k$-anonymity metric of privacy\\
\cline{2-5}
&~\cite{Mohammad2019Mobile}&Motion data&A multi-objective loss function involving an information-theoretic approach & Concealing user's private identity\\
\cline{2-5}
&\cite{Maximov2020CIAGAN}&Image and video&Conditional generative adversarial networks&Removing the identifying characteristics of faces and bodies for privacy\\
\hline\hline
\multirow{4}*{\shortstack{Dummy}}
&\cite{samarati1998protecting,Samarati2001Protecting,el2008protecting,li2007t}&Tabular dataset&Generating fake samples to hide real one& Realizing $k$-anonymity or similar metrics for privacy\\
\cline{2-5}
&\cite{Shaker2020Generalization}&Balance MIT-BIH arrhythmia dataset&Generative adversarial networks (GANs)& Generating high quality dummy samples for privacy\\
\hline\hline
\multirow{5}*{\shortstack{DP}}
&\cite{duchi2013local,Wang2017Locally,erlingsson2014rappor,fanti2016building}&Localized or tabular dataset & Using random response to perturb the value of local data & Achieving LDP for privacy \\
\cline{2-5}
&\cite{Balcan2012Distributed}&PAC-learning from distributed data&General upper and lower bounds for quantities such as the teaching-dimension& Achieving DP without incurring any additional communication penalty for privacy\\
\cline{2-5}
&\cite{Omobayode2020Efficient}& Communication
bandwidth limitation and security concerns of data upload&Training autoencoder, Transmitting latent vectors&Reducing the communications overhead and protecting the data of the end users\\
\hline\hline
\multirow{7}*{\shortstack{Encryption}}
&\cite{Junbeom2013Improving}&Enforcement of access policies, Support of policies updates&Defining their own access policies over user attributes and enforce the policies on the distributed data&Securely manage the data distributed \\
\cline{2-5}
&\cite{David2021Scalable}&Complete ML workflow by enabling the execution of a cooperative GD&Multiparty homomorphic encryption&Preserving data and model confidentiality with up to $N-1$ colluding parties\\
\cline{2-5}
&\cite{Xu2015Privacy}&Distributed training data, a large volume of the shared data portion. &Data locality property of Apache Hadoop architecture, a limited number of cryptographic operations&Achieving privacy-preservation with an affordable computation overhead\\
\hline\hline
\multirow{1}*{\shortstack{Others}}
&\cite{Zhang2015Secure}&A learner with a distributed set of nodes&Establishing a game-theoretic framework to capture the conflicting interests between the adversary and data processing units& Obtaining the network topology with a strong relation to the resiliency\\
\hline
\end{tabular}
\label{tab:defences_data_sharing}
\end{table*}

\subsubsection{Real Examples for Level-0 Distributed ML}
\begin{itemize}
    \item {RAPPOR}. Randomized aggregatable privacy-preserving ordinal response provides a privacy-preserving way to learn software statistics to better safeguard users¡¯ security, find bugs, and improve the overall user experience. Building on the concept of randomized response, RAPPOR enables learning statistics about the behavior of users¡¯ software while guaranteeing client privacy~\cite{erlingsson2014rappor}. The guarantees of differential privacy, which are widely accepted as the strongest form of privacy, have almost never been used in practice despite intense academic research. RAPPOR introduces a practical method to achieve those guarantees. In detail, the core of RAPPOR is a randomized response mechanism~\cite{warner1965randomized} for a user to answer a yes/no query to the record aggregator. A classic example is to collect statistics about a sensitive group, in which the aggregator asks each individual: ``Are you a doctor?" To answer this question, each individual tosses a coin, gives the true answer if it is a head, and a random yes/or answer otherwise. This randomized approach provides plausible deniability to the individuals. Meanwhile, it is shown to satisfy $\epsilon$-LDP, and the strength of privacy protection (i.e., $\epsilon$) can be controlled by using a biased coin. Based on the collected randomized answers, the aggregator estimates the percentage of users whose true answer is ``yes'' (resp. ``no''). RAPPOR allows the software to send reports that are effectively indistinguishable and are free of any unique identifiers. RAPPOR is currently an available implementation in Chrome, which learns statistics about how unwanted software is hijacking users¡¯ settings.
    \item {DP in the IOS system.} Apple has adopted and further developed local DP to enable Apple to learn about the user community while avoiding learning about individuals~\cite{2022Learning}. DP perturbs the information shared with random noise before it ever leaves the user¡¯s device such that Apple can never reproduce the raw data.  The power of additive noise that has been added can be reduced without exposing raw data from users by averaging out over large numbers of data points, and meaningful information emerges.
    DP is utilized as the first step of a system for data analysis that consists of robust privacy protections at every stage. The system is optional and developed to provide transparency to users. Device identifiers are removed from the data, and it is transmitted to Apple over an encrypted channel. The Apple analysis system ingests the differentially private contributions, dropping IP addresses and other metadata. The final stage is aggregation, where the private records are processed to compute the relevant statistics and the aggregate statistics are then shared with relevant Apple teams. Since both the ingestion and aggregation stages are performed in a restricted access environment, the raw data is not broadly accessible to the public.
\end{itemize}

\subsubsection{Brief Summary}
The guarantee of privacy and security in terms of data sharing models relies on the pre-processing of the raw data, such as perturbation, dummy, anonymization, and encryption. As shown in Fig.~\ref{fig:data-agg}, data pre-processing happens at the first stage of a ML task, and thus, these pre-processing techniques are usually harmful to the utility of systems or involved extra computations.
Therefore, it is more practical to select a proper mechanism to hide sensitive information from shared data while alleviating the negative influences on the system's utility.
\subsection{Level 1: Sharing Model}
\begin{figure*}
\centering
    \includegraphics[width=0.95\textwidth]{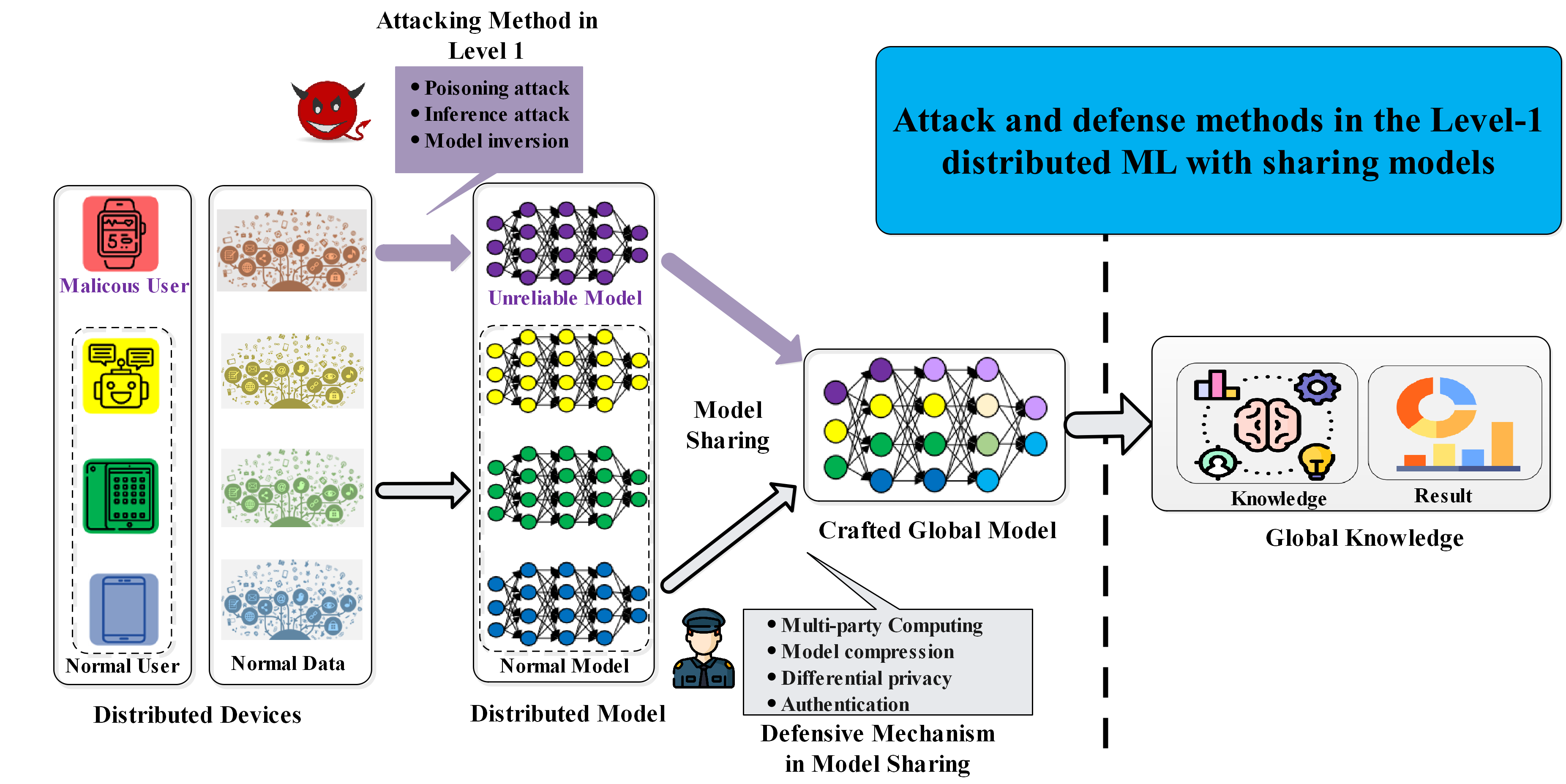}
    \caption{A breakout figure from Fig.~\ref{fig:DML}: an illustration of privacy and security issues in Level 1 distributed learning with sharing model.}
    \label{fig:model-agg}
\end{figure*}
\begin{table*}[hbt]
\label{Summary of attacks}
\caption{Taxonomy of attacks in Level-1 distributed ML with sharing models.}
\centering
\begin{tabular}{|m{2.5cm}<{\centering}||m{1.5cm}<{\centering}|m{4cm}<{\centering}|m{4cm}<{\centering}|m{4cm}<{\centering}|}
\hline
\textbf{Issue}& \textbf{Ref.}& \textbf{Attacker's knowledge}& \textbf{Learning Model}& \textbf{Effectiveness}\\
\hline\hline
\multirow{5}*{\shortstack{Model poisoning}}
&\cite{bagdasaryan2018backdoor}&Black-box&LSTM, ResNet&Manipulating the RL to achieve the designated purposes\\
\cline{2-5}
&\cite{bhagoji2018analyzing}&Black-box&CNN&Manipulating the RL to achieve the designated purposes\\
\cline{2-5}
&\cite{fang2020local}&White-box, Black-box&LR, CNN&Destroying the system performance\\
\hline\hline
\multirow{7}*{\shortstack{Inference attacks\\(Snooping attack)}}
&\cite{melis2019exploiting}&Black-box&CNN&Inferring certain sensitive characteristics of clients, such as locations and gender, etc.\\
\cline{2-5}
&\cite{Pan2019How}&Black-box access to the trained policy, access to the state space, the action space, the initial state distribution and the reward function&DQN, PG, PPO&Inferring certain sensitive characteristics of the training environment transition dynamics, such as dynamics coefficients, environment transition dynamics\\
\cline{2-5}
&\cite{Zhao2020Blackbox}&Black-box&DQN, A2C&Consistently predicting RL agents' future actions with high accuracy \\
\hline\hline
\multirow{4}*{\shortstack{Model inversion}}
&\cite{hitaj2017deep}&Black-box&CNN&Reconstructing raw training data\\
\cline{2-5}
&\cite{Wang2019Beyond}&Black-box&CNN&Reconstructing the actual training samples without affecting the standard training\\
\hline
\end{tabular}
\label{tab:attacks_model_sharing}
\end{table*}
\begin{table*}[hbt]
\centering
\caption{Taxonomy of defenses in Level-1 distributed ML with sharing models.}
\begin{tabular}{|m{2cm}<{\centering}||m{1.5cm}<{\centering}|m{6cm}<{\centering}|m{2.5cm}<{\centering}|m{4cm}<{\centering}|}
\hline
\textbf{Method}& \textbf{Ref.} &\textbf{Description}& \textbf{Key Challenges}&\textbf{Effectiveness}\\
\hline\hline
\multirow{1}*{\shortstack{DP}}
&\cite{Jason2018Minimax,Huang2020ADMM,abadi2016deep,Robin2017Differentially, Wei2020Federated,Xiao2021Arbitrarily,Ye2020Differentially}& Introducing a level of uncertainty into the released model sufficient to mask the contribution of any individual user&Finding a balance between the training performance and privacy level&Low complexity in preserving privacy\\
\hline\hline
\multirow{1}*{\shortstack{Model\\compression}}
&\cite{Li2019Privacy,Venkata2021vqSGD}&Encoding local models before transferring them to the server&Measuring the effect on the privacy and reduce the negative effect on the training performance&Low complexity and high communication efficiency\\
\hline\hline
\multirow{1}*{\shortstack{HE}}
&\cite{Phong2018Privacy,Xu2020VerifyNet}& Mathematical operations applied on an encrypted message result in the same mathematical operation being applied to the original message&Increasing computation complexity and transmission bits&Strongly effective in security\\
\hline\hline
\multirow{1}*{\shortstack{Secure MPC}}
&\cite{bonawitz2017practical}&Allowing two or more participants to jointly compute functions over their collective data without disclosing any sensitive information& Lack of a common protocol for various tasks& A lower complexity than HE and a higher security than DP\\
\hline\hline
\multirow{1}*{\shortstack{ Statistical\\analysis}}
&\cite{Mu2019Byzantine,Fu2019attack}&Detecting and filtering the outliers
based on the statistical information, e.g., Euclidean distance and principle component &Destroying the training performance especially in the non-i.i.d. setting&Low complexity to detect outliers\\
\hline\hline
\multirow{1}*{\shortstack{Pretest on \\Auxiliary Datasets}}
&\cite{Zhao2020shielding,Zhao2020PDGAN}&Calculating the accuracy score for all local models and reducing the effect of low-quality ones&Performance governed by the quality of auxiliary datasets& Directly detecting malicious users with sensitive datasets\\
\hline\hline
\multirow{3}*{\shortstack{Authentication}}
&\cite{Wang2007Formal}&Using trust composition for determining the trust and reputation values for unknown agents&Relying on the trust transfer and vulnerable to the collusion&Low complexity in security\\
\cline{2-5}
&\cite{Danilov2018Towards,Strobel2018Managing,Calvaresi2019Explainable}&Combining blockchain technology and reaching an agreement by a group of agents&Vulnerable to the $51\%$ attack&Guaranteeing fairness in integrity\\
\hline\hline
\multirow{1}*{\shortstack{Authorization}}&\cite{Xiao2007An,Ahmad2012A,Bosse2016Mobile}&Constructing capability-based access and different agent privilege levels&Formulating corresponding authorization standards for differential privilege levels&Guaranteeing the quality of participants\\
\hline
\end{tabular}
\label{tab:defences_model_sharing}
\end{table*}

In model sharing systems, all distributed nodes need to share their training models with the central server or other participants.
Via the interaction between independent data training and local model aggregation, model sharing systems can capture a required learning model over data that resides at the associated nodes.
\subsubsection{Threat Models}
Although data is not required to upload in model sharing systems, private information can still be divulged by analyzing uploaded model parameters, e.g., weights trained in deep neural networks.
Moreover, adversarial participants may degrade or even destroy the training systems by uploading unreliable models.
Attacks can be carried out by the following three aspects.
\begin{itemize}
\item[$\bullet$] \textbf{Insiders vs. outsiders.}
Insider attacks include those launched by the server and the participants in the model sharing systems.
Outsider attacks include those launched by the eavesdroppers in the wireless transmission environment between participants and the server, and by users of the final model when it is deployed as a service.
Insider attacks are generally stronger than outsider attacks, as it strictly enhances the capability of the adversary.
\item[$\bullet$] \textbf{Semi-honest vs. malicious.}
Under the semi-honest setting, adversaries are considered passive or honest but curious.
They try to learn the private states of other participants without deviating from the model sharing protocol.
The passive adversaries are assumed to only observe the aggregated or averaged gradient, but not the training data or gradient from other honest participants.
Under the malicious setting, an active, or malicious adversary tries to learn the private states of honest participants and deviates arbitrarily
from the model sharing protocol by modifying, re-playing, or removing messages.
This strong adversary model allows the adversary to conduct particularly devastating attacks.
\item[$\bullet$] \textbf{Poisoning vs. inference.}
Attacks at the poisoning phase attempt to learn, influence, or corrupt the model sharing itself.
During the poisoning phase, the attacker can run data poisoning attacks to compromise the integrity of training dataset collection, or launch model poisoning attacks to compromise the integrity of the learning process.
The attacker can also launch a range of inference attacks on an individual participant's update or on the aggregation of updates from all participants.
\end{itemize}
\subsubsection{Taxonomy of Attacks}
Attacks to model sharing models can be categorized as poisoning attacks, inference attacks, and model inversion based on their various goals as shown in Table~\ref{tab:attacks_model_sharing}. We also summarize them as follows.
\begin{itemize}
\item[$\bullet$] \textbf{Poisoning attack.} Compromised clients by attackers always have opportunities to poison the global model in model sharing systems, in which local models are continuously updated by clients throughout their deployments.
Moreover, the existence of compromised clients may induce further security issues such as bugs in pre-processing pipelines, noisy training labels, as well as explicit attacks that target training and deployment pipelines~\cite{kairouz2021advances}.
In order to destroy ML models, poisoning attackers may control part of clients and manipulate their outputs sent to the server.
For example, the compromised clients can upload noisy and reversed models to the server at each communication round~\cite{Huang2011adver,Fu2019attack}, which has the advantage of low complexity to mount attacks.
Other attackers may manipulate the outputs of the compromised clients carefully to achieve the evasion of defenses and downgrade the performance of ML models.
Furthermore, the authors in \cite{fang2020local,baruch2019little} have formulated the local model poisoning attack as optimization problems, and then apply this attack against recent Byzantine-robust FL methods.
In this way, attackers can improve the success rate of attacks, dominate the cluster and change the judgment boundary of the global model, or make the global model deviate from the right direction.
Besides, attackers may hope to craft the ML model to minimize this specific objective function instead of destroying it.
Via using multiple local triggers and model-dependent triggers (i.e., generated based on local models of attackers), the collusive attackers can conduct backdoor attacks successfully~\cite{Gong2022Coordinated}.
Bagdasaryan~\emph{et al.} in~\cite{bagdasaryan2018backdoor} have developed and evaluated a generic constrain-and-scale technique that incorporates the evasion of defenses into the attacker's loss function during training.
The work in~\cite{bhagoji2018analyzing} has explored the threat of model poisoning attacks on FL initiated by a single, non-colluding malicious client where the adversarial objective is to cause the model to misclassify a set of chosen inputs with high confidence.
\item[$\bullet$] \textbf{Inference attack.}
The work in~\cite{baruch2019little} has presented a new attack paradigm, in which a malicious opponent may interfere with or backdoor the process of distributed learning by applying limited changes to the uploaded parameters.
The work in~\cite{bagdasaryan2018backdoor} has proposed a new model-replacement method that demonstrated its efficacy on poisoning models of standard FL tasks.
Inferring privacy information about clients for attackers is also possibly achievable in ML models.
A generic attacking framework mGAN-AI that incorporates a multi-task GAN has been proposed in~\cite{Luo2020Feature}, which conducted novel discrimination on client identity, achieving attack to clients' privacy, i.e., discriminating a participating party's feature values, such as category, reality, and client identity.
\item[$\bullet$] \textbf{Model inversion.}
By casting the model inversion task as an optimization problem, which finds the input that maximizes the returned confidence, the work in ~\cite{fredrikson2015model} has recovered recognizable images of people's faces given only their names and accesses to the ML model.
In order to identify the presence of an individual's data, an attack model trained by the shadow training technique has been designed, and can successfully distinguish the target model's outputs on members versus non-members of its training dataset~\cite{shokri2015privacy}.
\end{itemize}

Specifically, in distributed reinforcement learning (DRL) systems, there has been literature available on security vulnerabilities.
We provide many characteristics of an adversary's capabilities and goals that can be studied as follows.
First, we divide attacks based on what components in an MDP the attacker chooses to attack: the agent's observations, actions, and environment (transition) dynamics.
Then, we discuss the practical scenarios where attacks happen on these components.
\begin{itemize}
\item[$\bullet$]\textbf{Observations.} Existing work on attacking DRL systems with adversarial perturbations focuses on perturbing an agent's observations, i.e., states and rewards, that are communicated between the agent and the environment.
This is the most appealing place to start, with seminal results already suggesting that recognition systems are vulnerable to adversarial examples~\cite{Sandy2017Adversarial,chen2017Tactics, Behzadan2017Vulnerability,Alessio2019Optimal,Ma2019Policy,Panagiota2019TrojDRL,Zhang2020Adaptive, Huai2020Malicious,Sun2020Stealthy,Rakhsha2020Policy}.
Sandy~\emph{et al.}~\cite{Sandy2017Adversarial} have first shown that adversarial attacks are also effective when targeting neural network policies in RL adversarial examples.
Based on this technique, part of the works enhance adversarial examples to attack DRL.
To improve the attack efficiency, the strategically-timed attack~\cite{chen2017Tactics}, consuming a small subset of time steps in an episode, has been explored.
Via stamping a small percentage of inputs of the policy network with the Trojan trigger and manipulating the associated rewards, the work in~\cite{Panagiota2019TrojDRL} has proposed the TrojDRL attack, which can deteriorate drastically the policy network in both targeted and untargeted settings.
Another fancy idea for a reward-poisoning attack is to design an adaptive disturbing strategy~\cite{Zhang2020Adaptive}, where the infinity norm constraint is adjusted on the DRL agent's learning process at different time steps.
For the theoretical analysis, two standard victims with adversarial observations, i.e., tabular certainty equivalence learner in reinforcement learning and linear quadratic regulator in control have been analyzed in a convex optimization problem on which global optimality and the attack feasibility and attack cost have been provided~\cite{Ma2019Policy}.
In addition, the effectiveness of a universal adversarial attack against DRL interpretations (i.e., UADRLI) has been verified by the theoretical analysis~\cite{Huai2020Malicious}, from which the attacker can add the crafted universal perturbation uniformly to the environment states in a maximum number of steps to incur minimal damage.
In order to stealthily attack the DRL agents, the work in~\cite{Sun2020Stealthy} has injected adversarial samples in a minimal set of critical moments while causing the most severe damage to the agent.
Another work in~\cite{Rakhsha2020Policy} has formulated an optimization framework in a stealthy manner to find an optimal attack for different measures of attack cost and solved it with an offline and online setting.
\item[$\bullet$]\textbf{Actions.}
Attacks applied on the action space usually  {\color{blue}aim} to minimize the expected return or lure the agent to a designated state, e.g., the action outputs can be modified by installing some hardware virus in the actuator executing process.
This can be realistic in certain robotic control tasks where the control center sends some control signals to the actuator. A vulnerability in the implementation, i.e., the vulnerability in the blue-tooth signal transmission, may allow an attacker to modify that signal~\cite {Lonzetta2018Security}.
A training policy network to learn the attack has been developed, which treats the environment and the original policy together as a new environment, and views attacks as actions~\cite{Xiao2019Characterizing}.
However, existing works only concentrate on the white-box scenario, i.e., knowing the victim's learning process and observations, which is not practical and inaccessible to attackers.
\item[$\bullet$]\textbf{Environment Dynamics.}
The environment (transition) dynamics can be defined as a probability mapping from state-action pairs to states, which is governed by the environmental conditions.
For attacks applied on the environment dynamics, an attacker may infer environment dynamics\cite{Pan2019How} or perturb a DRL system's environment dynamics to make an agent fail in a specific way~\cite{Xiao2019Characterizing,Adam2020Adversarial,Zhong2020Adversarial,Rakhsha2020Policy}.
In the autonomous driving case, the attacker can change the material surface characteristic of the road such that the policy trained in one environment will fail in the perturbed environment.
In a robot control task, the attacker can change the robot's mass distribution so that the robot may lose balance when executing its original policy because it has not been trained in that case.
\end{itemize}

Then, we categorize these attacks based on what knowledge the attacker needs.
Broadly, this breaks attacks down into the already recognized white-box attacks, where the attacker has full knowledge of the DRL system, and black-box attacks, where the attacker has less or no knowledge.
\begin{itemize}
\item[$\bullet$]\textbf{White-Box.}
If the adversary attacks the DRL system with the capability of accessing the architecture, weight parameters of the policy and Q networks, and querying the network, we can call it a white-box attack.
Clearly, the attacker can formulate an optimization framework for the white-box setting and derive the optimal adversarial perturbation~\cite{Sandy2017Adversarial,Huai2020Malicious}.
Moreover, via the theoretical analysis of the attack feasibility and cost, the adversary can further decrease the efficiency and stealth of the learning~\cite{Ma2019Policy,Xiao2019Characterizing}.
However, this setting is inaccessible for the adversary in most practical scenarios.

\item[$\bullet$]\textbf{Black-Box.}
In general, the trained RL models are kept private to avoid easy attacks by certain secure access control mechanisms.
Therefore, the attacker cannot fully acknowledge the weight parameters of the policy network and Q networks, and may or may not have access to query the policy network.
In this case, the attacker can train a surrogate policy to imitate the victim policy, and then use a white-box method on the surrogate policy to generate a perturbation and applies that perturbation to the victim policy~\cite{Xiao2019Characterizing}.
The finite difference (FD) method~\cite{Bhagoji_2018_ECCV} in attacking classification models can be utilized to estimate the gradient on the input observations, and then perform gradient descent to generate perturbations on the input observations~\cite{Xiao2019Characterizing}.
In this black-box setting, the adversary becomes difficult to perturb a DRL system and needs to estimate the victim's information with large computation costs, such as policies and observations.
\end{itemize}

Based on the adversary's objective, adversarial attacks are divided into two types: poisoning attacks and snooping attacks.
\begin{itemize}
\item[$\bullet$]\textbf{Poisoning Attack.}
In particular, for poisoning attacks, there are at least two dimensions to potential attacks against learning systems as untargeted attacks~\cite{Sandy2017Adversarial} and targeted (induction) attacks~\cite{Behzadan2017Vulnerability}.
In untargeted attacks, attackers focus on the integrity and availability of the DRL system, i.e., minimizing the expected return (cumulative rewards).
Specifically, the work ~\cite{Sandy2017Adversarial} has shown existing adversarial example crafting techniques can be used to significantly degrade the test-time performance of trained policies.
However, in terms of defensive mechanisms, the attacker may control time steps~\cite{Sun2020Stealthy} or solve an optimization framework in a stealthy manner~\cite{Huai2020Malicious}.
Another attack of this category aims at maliciously luring an agent to a designated state more than decreasing the cumulative rewards~\cite{Behzadan2017Vulnerability}.
Via combining a generative model and a planning algorithm, the generative model predicts the future states and the planning algorithm generates a preferred sequence of actions for luring the agent~\cite{chen2017Tactics}.
Similar to untargeted attacks, by solving an optimization framework in a stealthy manner~\cite{Rakhsha2020Policy}, the attacker can easily succeed in teaching any target policy.
\item[$\bullet$]\textbf{Snooping Attack.}
Different from poisoning attacks, the attacker only aims to eavesdrop on environment dynamics, the action and reward signals being exchanged between the agent and the environment.
If the adversary can train a surrogate DRL model that closely resembles the target agent~\cite{Pan2019How,Zhao2020Blackbox}, the desired information can be estimated by this model.
Furthermore, the adversary only needs to train a proxy model to maximize reward, and adversarial examples crafted to fool the proxy will also fool the agent~\cite{Inkawhich2020Snooping}.
We can note that the snooping attacks can still launch devastating attacks against the target agent by training proxy models on related tasks, and leveraging the transfer-ability of adversarial examples.
\end{itemize}
\subsubsection{Taxonomy of Defences}
Defensive mechanisms found in multiple works of literature are grouped by their underlying defensive strategies as shown in Table~\ref{tab:defences_model_sharing}. We will discuss various defenses in model sharing frameworks as follows.
\begin{itemize}
\item[$\bullet$]\textbf{DP.}
DP tackles the privacy leakage about the single data change in a dataset when some information from the dataset is publicly available and is widely used due to its strong theoretical guarantees~\cite{Liu2022Projected}.
Common DP mechanisms will add an independent random noise component to access data, i.e., the shared models in this level, to provide privacy.
DP preserving distributed learning systems have been studied from various paradigms, such as distributed principal component analysis (PCA)~\cite{Jason2018Minimax}, distributed ADMM~\cite{Huang2020ADMM}, distributed SGD~\cite{abadi2016deep}, FL~\cite{Robin2017Differentially,Wei2020Federated} and multi-agent reinforcement learning~\cite{Xiao2021Arbitrarily,Ye2020Differentially}.
In order to provide fine-tuned control over the trade-off between the estimation accuracy and privacy preservation, a distributed privacy-preserving sparse PCA (DPS-PCA) algorithm that generates a min-max optimal sparse PCA estimator under DP constraints has been proposed in~\cite{Jason2018Minimax}.
Similarly, for distributed ADMM, distributed SGD, FL, and multi-agent reinforcement learning systems, all related works focus on improving the utility-privacy trade-off via two aspects as follows:
a) analyzing the learning performance with a DP constraint and then optimizing system parameters;
b) enhancing the DP mechanism by obtaining tighter estimates of the overall privacy loss.
\item[$\bullet$]\textbf{Model compression.}
Model compression techniques for distributed SGD and FL systems, e.g., sketches, can achieve provable privacy benefits~\cite{Li2019Privacy,Venkata2021vqSGD}.
Therefore, a novel sketch-based framework (DiffSketch) for distributed learning has been proposed, improving absolute test accuracy while
offering a certain privacy guarantee and communication compression.
Moreover, the work in~\cite{Venkata2021vqSGD} has presented a family of vector quantization schemes, termed Vector-Quantized
Stochastic Gradient Descent (VQSGD), provides an asymptotic reduction in the communication cost and automatic privacy guarantees.
\item[$\bullet$]\textbf{Encryption.}
Encryption, e.g., HE~\cite{Phong2018Privacy} and MPC~\cite{bonawitz2017practical}, is also adopted to protect user data privacy through parameter exchange under the well-designed mechanism during ML.
A novel deep learning system~\cite{Phong2018Privacy}, bridging asynchronous SGD and cryptography, has been proposed to protect gradients over the honest-but-curious cloud server, using additively homomorphic encryption, where all gradients are encrypted and stored on the cloud server.
To verify whether the cloud server is operating correctly, VerifyNet~\cite{Xu2020VerifyNet} has been proposed to guarantee the confidentiality of users' local gradients via a double-masking protocol in FL, where the cloud server is required to provide proof of the correctness of its aggregated results to each user.
\item[$\bullet$]\textbf{MPC.}
The work in~\cite{bonawitz2017practical} has outlined an approach to advancing privacy-preserving ML by leveraging MPC to compute sums of model parameter updates from individual users' devices in a secure manner. The problem of computing a multiparty sum where no party reveals its updates to the aggregator is referred to  {\color{blue}as} secure aggregation. Via encoding local models into multiple secret shares in the first round, and then splitting each share into a public share and a private share, the work in ~\cite{Zhang2022Augmented} can provide stronger protections for the security and privacy of the training data. MPC integrates the encryption technology and interactive protocols, aiming to make the receiver keep away from sensitive information and obtain the necessary messages ~\cite{Renuga2020Two,Ekanut2021Partially,Mou2021A,Shamsabadi2020PrivEdge}.
\item[$\bullet$]\textbf{Statistical analysis.}
The work in~\cite{Mu2019Byzantine} has proposed a robust aggregation rule, called adaptive federated averaging, that detects and discards bad or malicious local model updates based on a hidden Markov model.
To tackle adversarial attacks in the FL aggregation process, the work in~\cite{Fu2019attack} presented a novel aggregation algorithm with the residual-based re-weighting method, in which the weights for the average of all local models are estimated robustly. Via controlling the global model smoothness based on clipping and smoothing on model parameters, a sample-wise robustness certification FL framework has been proposed, which can train certifiably robust FL models against backdoors~\cite{Xie2021CRFL}. Most of the defenses for FL aim to explore the latent model exception, such as similarities between malicious and benign clients, and then lessen the influence of these exceptional models~\cite {Cao2023FedRecover,Cao2022FLCert,Zhang2022FLDetector,Farnaz2022RobustFed}.
\item[$\bullet$]\textbf{Pretest on auxiliary datasets}
For detecting poisoned updates in collaborative learning~\cite{Zhao2020shielding}, the results of client-side cross-validation were applied for adjusting the weights of the updates when performing aggregation, where each update is evaluated over other clients' local data.
The work in~\cite{Zhao2020shielding} considered the existence of unreliable participants and used the auxiliary validation data to compute a utility score for each participant to reduce the impact of these participants.
The work in~\cite{Zhao2020PDGAN} has proposed a novel poisoning defense method in FL, in which the participant whose accuracy is lower than a predefined threshold will be identified as an attacker, and the corresponding model parameters will be removed from the training procedure in this iteration.
\item[$\bullet$]\textbf{Authentication and access control.}
The key question in considering security in a MARL consists of increasing the confidence that all parties involved in the system (agents, platforms, and users) will behave correctly, and this can be achieved through the authentication of these parties.
The identification of the parties can make up a system and possibly establish an agent-trust relationship.
Thus, how to design efficient identity certification mechanisms to uniquely authenticate known and trusted users and agents in the system has drawn heated attention.
A domain-independent and reusable MARL infrastructure has been developed in~\cite{Petr2003Communication}, in which the system uses a certification authority (CA) and ensures full cooperation of secured agents and already existing (unsecured) agents.
The work in~\cite{Wang2007Formal} has introduced a method called trust composition, which combines several trust values from different agents.
We can note that the trust composition can play a critical role in determining the trust and reputation values of unknown agents since it is impractical for an agent to get complete knowledge about other agents.
A work called PTF (Personalized Trust Framework) has been proposed to establish a trust/reputation model for each application with personalized requirements~\cite{Huynh2009A}.
Naturally, the idea of using blockchain technology to solve security problems in multi-robot systems was discussed in~\cite{Danilov2018Towards}.
The work in~\cite{Danilov2018Towards} stated that combining peer-to-peer networks with cryptographic algorithms allows reaching an agreement by a group of agents (with the following recording this agreement in a verifiable manner) without the need for a controlling authority.
Thus, blockchain-based innovations can provide a breakthrough in MARL applications.
The work in~\cite{Strobel2018Managing} has developed an approach to using decentralized programs based on smart contracts to create secure swarm coordination mechanisms, as well as for identifying and eliminating Byzantine swarm members through collective decision-making.
The work in~\cite{Calvaresi2019Explainable} has proposed an approach combining blockchain technology and explainability supporting the decision-making process of MARL, in which blockchain technology offers a decentralized authentication mechanism capable of ensuring trust and reputation management.

\item[$\bullet$]\textbf{Authorization and trust model.}
Combined with authentication, authorization is used to restrict the actions that an agent can perform in a system, and control the access to resources by these agents.
Sensitive information about principals is transferred online even across the Internet and is stored in local and remote machines.
Without appropriate protection mechanisms, a potential attacker can easily obtain information about principals without their consent.
In the context of authorization mechanisms, the algorithm proposed in~\cite{Xiao2007An} is designed to solve the problem of systems that are constantly changing.
The main goal is to build a flexible and adaptive security policy management capable to configure itself and reflect the actual needs of the system.
According to the authors, a system is not safe if a security model is developed but never managed afterward.
Security of the proposed system in~\cite{Ahmad2012A} has been further explored in the form of authorization and encryption of the data by introducing an authorization layer between the user and the system that will be responsible for providing access to the legitimate users of the system only.
The work in~\cite{Bosse2016Mobile} has ensured agent authorization and platform security with capability-based access and different agent privilege levels, in which the agent behavior is modeled with an activity transition graph (ATG) and implemented entirely in JavaScript with a restricted and encapsulated access to the platform API (AgentJS).
\end{itemize}
\subsubsection {Real Examples for Level-1 Distributed ML}

\begin{itemize}
    \item Electronic Medical Records (EMR)~\cite{2022AI}. The use of information and network technologies in the healthcare field inevitably produces EMR, which is a necessary trend for the modernization of medical records in hospitals. The initial adoption of EMR in clinical practice has vastly improved the efficiency and quality of health care provided by hospitals. Empowered by algorithm technologies and data reconstruction, BaseBit~\cite{2022AI} has constructed a robust and comprehensive knowledge base system and has a series of intelligent models with excellent abilities of expression. In various applications centered around electronic medical records, the proposed models effectively improve the abilities such as automatic medical record writing, overall quality control, cost monitoring systems for single diseases, early warning for infectious diseases, prompt for critical illnesses, clinical decision-making assistance for rare diseases, enabling hierarchical diagnosis and treatments.
\end{itemize}
\subsubsection{Brief Summary}
As shown in Fig.~\ref{fig:model-agg}, although due to the local training process, the raw data of each participant will not be exposed to the curious server or external attackers, defensive mechanisms are also necessary because of the existing possibility of feature inference and data reconstruction from models sharing, in addition to the model poisoning paradigm. Traditional HE and DP are proven beneficial to privacy preservation but lead to low efficiency or damaged utility.
Therefore, the quantitative analysis of the relationship between the sensitive feature and the published model is imperative.

\begin{table*}[hbt]
\caption{Taxonomy of attacks in Level-2 distributed ML with sharing knowledge.}
\centering
\begin{tabular}{|m{2.5cm}<{\centering}||m{1.5cm}<{\centering}|m{3.5cm}<{\centering}|m{4cm}<{\centering}|m{4cm}<{\centering}|}
\hline
\textbf{Method}& \textbf{Ref.}& \textbf{Attacker's knowledge}& \textbf{Learning Model}& \textbf{Effectiveness}\\
\hline\hline
\multirow{1}*{\shortstack{Label leakage}}
&\cite{Oscar2021Label}&Black box&Split learning &Revealing the ground-truth labels from the participants\\
\hline
\multirow{1}*{\shortstack{Feature inference}}
&\cite{Luo2020Feature}&Black box&Vertical FL&Inferring the feature values of new samples belong
to the passive parties successfully\\
\hline
\multirow{3}*{\shortstack{Data reconstruction}}
&\cite{Sharif2020Can}&Black box&Split learning & Activated output
after two and three convolutional layers can be used to reconstruct the raw data \\
\cline{2-5}
&\cite{Weng2020Privacy}&Black box&Vertical FL& Stealing partial raw training data successfully\\
\hline
\end{tabular}
\label{tab:attacks_knowledge_sharing}
\end{table*}
\subsection{Level 2: Sharing Knowledge}
\begin{figure*}
\centering
    \includegraphics[width=0.95\textwidth]{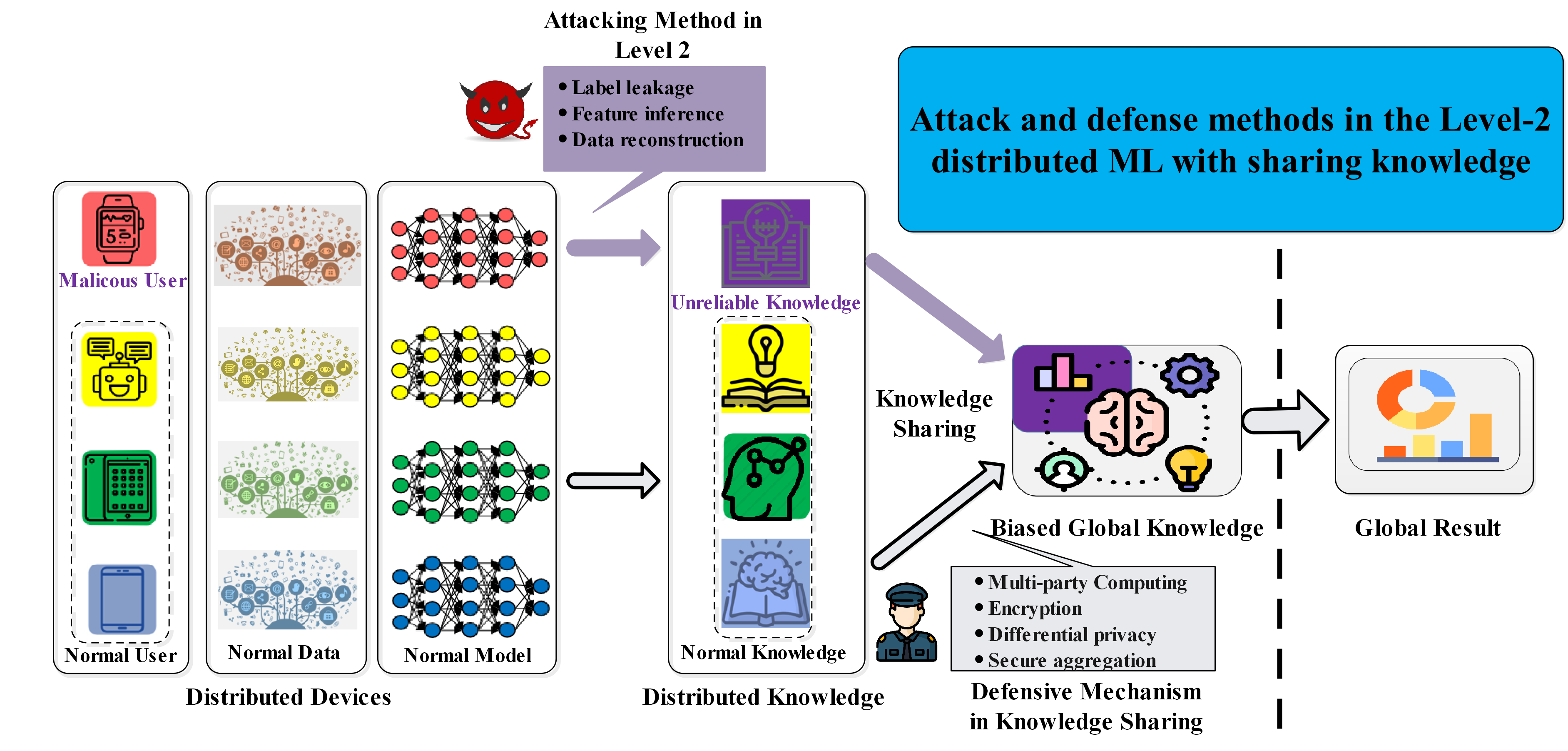}
    \caption{A breakout figure from Fig.~\ref{fig:DML}: an illustration of privacy and security issues in Level 2 distributed learning with sharing knowledge.}
    \label{fig:knowledge-agg}
\end{figure*}
Recent configurations that rely on knowledge sharing techniques can be summarized as split learning~\cite{vepakomma2018split}, vertical FL~\cite{yang2019federated}, and distillation-based FL~\cite{Sohei2021Distillation}.
Split learning allows multiple clients to hold different modalities of vertically partitioned data and learn partial models up to a certain layer (the so-called cut layer).
Then the outputs at the cut layer from all clients are then concatenated and sent to the server that trains the rest of the model.
In vertical FL, participants hold the same set of samples but with disjoint features and only one participant owns the labels, which need to combine split NNs and privacy-preserving techniques~\cite{Daniele2021PyVertical}.
Distillation-based FL~\cite{Sohei2021Distillation,Sun2021Federated,He2020Group} exchanges model outputs instead of model parameters, where the communication overhead cannot scale up according to the model
size and has been proven to satisfy the DP guarantee.

\subsubsection{Threat Models}
In knowledge sharing paradigms, adversarial participants or eavesdroppers still possibly exist.
The adversarial participants can be categorized into two kinds: a) honest-but-curious (semi-honest) participants, who do not deviate from the defined learning protocol, but attempt to infer private training data from the legitimately received information; b) malicious participants, who may deviate from the defined learning protocol, and destroy this training task or inject Trojans to the training model.
\subsubsection{Taxonomy of Attacks} Existing attacks on knowledge sharing paradigms can be mainly categorized as label leakage, feature inference, and data reconstruction as shown in Table~\ref{tab:attacks_knowledge_sharing}. Then, we discuss existing attacks as follows.
\begin{itemize}
\item[$\bullet$] \textbf{Label leakage.}
The labels in distributed learning frameworks might be highly sensitive, e.g., whether a person has a certain kind of disease.
However, the bottom model structure and the gradient update mechanism of VFL or split learning can be exploited by a malicious participant to gain the power to infer the privately owned labels~\cite{Ege2022UnSplit}.
Worse still, by abusing the bottom model, he/she can even infer labels beyond the training dataset~\cite{Fu2022Label}.
The work in~\cite{Oscar2021Label} first made an attempt at a norm attack that uses the norm of the communicated gradients between the parties, and it can largely reveal the ground-truth labels from participants.
The adversary (either clients or servers) can accurately retrieve the private labels by collecting the exchanged gradients and smashed data~\cite{Liu2022Clustering}. Thus, it is necessary to make gradients from samples with different labels similar.
\item[$\bullet$] \textbf{Feature inference.} Through analysis, the work in~\cite{Ye2022Feature,Qiu2022Your} demonstrated that, unless the feature dimension is exceedingly large, it remains feasible, both theoretically and practically, to launch a reconstruction attack with an efficient search-based algorithm that prevails over current feature protection techniques.
In this paper, the authors have performed the first systematic study on relation inference attacks to reveal VFL's risk of leaking samples' relations. Specifically, the adversary is assumed to be a semi-honest participant. Then,
according to the adversary's knowledge level, the work ~\cite{Qiu2022Your} formulated three kinds of attacks based on different intermediate representations and revealed VFL's risk of leaking samples' relations.
Luo \emph{et al.}~\cite{Luo2020Feature} considered the most stringent setting that the active party (i.e., the adversary) only controls the trained vertical FL model and the model predictions, and then observed that those model predictions will leak a lot of information about features by learning the correlations between the adversary's and the attacking target's features.
\item[$\bullet$] \textbf{Data reconstruction.}
The work in~\cite{Sharif2020Can} has provided the leakage analysis framework via three empirical and numerical metrics
(distance correlation and dynamic
time warping) indicating that the activated outputs after two and more convolutional layers can be used to reconstruct the raw data, i.e., sharing the intermediate
activation from these layers may result in severe privacy leakage.
In vertical FL, two simple yet effective attacks, reverse multiplication attack and reverse sum attack, have been proposed to steal the raw training data of the target participant~\cite{Weng2020Privacy}.
Though not completely equivalent to the raw data, these stolen partial
orders can be further used to train an alternative model which is
as effective as the one trained on the raw data~\cite{Xiao2021}.
\end{itemize}
\subsubsection{Taxonomy of Defences} Defensive mechanisms found in multiple works of literature are grouped by their underlying defensive strategy as shown in Table~\ref{tab:defences_knowledge_sharing}. Hence, we will discuss various defenses in model sharing frameworks as follows.
\begin{itemize}
\item[$\bullet$] \textbf{DP.}
The work in~\cite{pathak2010multiparty} has proposed a privacy-preserving protocol for composing a differentially private aggregate classifier using local classifiers from different parties.
In order to overcome the effects of the proposed information inference attacks~\cite{Sharif2020Can}, DP has been proven helpful in reducing privacy leakage but leading to a significant drop in model accuracy.
\item[$\bullet$] \textbf{MPC.} The work in~\cite{Wu2020Privacy} has proposed a novel solution for privacy-preserving vertical decision tree training and prediction, termed Pivot, ensuring that no intermediate information is disclosed other than necessary releases (i.e., the final tree model and the prediction output).
\item[$\bullet$] \textbf{Encryption.} A novel privacy-preserving architecture has been proposed in~\cite{Zhang2020Additively}, which can collaboratively train a deep learning model efficiently while preserving
the privacy of each party's data via the HE technique.
The work in~\cite{Zhang2020Additively} has explored a lossless privacy-preserving tree-boosting system known as SecureBoost by using the additive HE scheme.
\item[$\bullet$] \textbf{Secure aggregation.} The work in~\cite{Tian2020FederBoost} has proposed the vertical FederBoost which runs the gradient boosting decision tree (GBDT) training algorithm in exactly the same way as the centralized learning.
Via further utilizing packetization and DP, this algorithm can protect the order of samples:
participants partition the sorted samples of a feature into buckets, which only reveals the order of the buckets and add differentially private noise to each bucket.
\item[$\bullet$] \textbf{Others.} The work in~\cite{Ang2020TIPRDC} has presented TIPRDC to learn a feature extractor that can hide the private information from the intermediate representations using an adversarial training process while maximally retaining the original information embedded in the raw data to accomplish unknown learning tasks.
In~\cite{Sharif2020Can}, adding more hidden layers to the client side was proven helpful in reducing privacy leakage, but increasing the number of layers seems ineffective with the most highly correlated
channels.
In order to relieve the negative impact of random perturbation preserving techniques on the learned model's predictive performance, the work in~\cite{Oscar2021Label} has introduced an improved way to add Gaussian noise by making the expected norm of the positive and negative gradients in a mini-batch equal (un-distinguishable).
\end{itemize}

\subsubsection{Real Examples for Level-2 Distributed ML}
\begin{itemize}
    \item FATE. An open-source project, named FATE, provides a secure computing framework to support the federated AI ecosystem~\cite{Yang2019Federated1}, led by Webank¡¯s AI Department. It can enable big data collaboration without privacy leakage by implementing multiple secure computation protocols, such as DP, HE, and so on. FATE accesses out-of-box usability and excellent operational performance with a modular modeling pipeline, explicit visual interface, and flexible scheduling system~\cite{2022Overview}. eHi Car Services, a national chain car rental brand, and WeBank jointly announced a deep strategic partnership, announcing that the two sides will carry out multi-scene and multi-dimensional innovation cooperation in car travel, member services, finance and insurance, blockchain technology, and other fields. eHi Car Services uses federal transfer learning, AI face authentication technology, payment technology, and other fin-techs to deeply integrate into the car rental service process for the purpose of optimizing and improving user experience, and combines the car rental scene with the bank's big data risk control system, so as to provide a new way of travel and life for the young and long-term rental customers.
\end{itemize}

\subsubsection{Brief Summary}
As shown in Fig.~\ref{fig:knowledge-agg}, split learning, vertical FL, and distillation-Based FL are the classical knowledge sharing systems, in which the knowledge can be viewed as the  {\color{blue}partial} processing result to meet the requirement of the system learning.
It is also challenging for knowledge sharing systems to hide sensitive information from the shared knowledge.
\begin{table*}[hbt]
\caption{Taxonomy of defences in Level-2 distributed ML with sharing knowledge.}
\centering
\begin{tabular}{|m{2.5cm}<{\centering}||m{1.5cm}<{\centering}|m{3.5cm}<{\centering}|m{4cm}<{\centering}|m{4cm}<{\centering}|}
\hline
\textbf{Method}& \textbf{Ref.}& \textbf{Use case}& \textbf{Key idea}& \textbf{Effectiveness}\\
\hline\hline
\multirow{4}*{\shortstack{DP}}
&\cite{pathak2010multiparty}&Deriving aggregate information without revealing information about individual data instances&Differentially private aggregate in a multi-party setting&DP analysis on the perturbed aggregate classifier\\
\cline{2-5}
&\cite{Sharif2020Can}&Against DCM and DTWM attacks in split learning&Laplace mechanism on the split layer activation&Strong DP level ($\epsilon=1$) works but degrading the classification accuracy\\
\hline\hline
\multirow{1}*{\shortstack{MPC}}
&\cite{Wu2020Privacy}&Vertical decision tree training, random forest (RF), and gradient boosting
decision tree (GBDT) &A hybrid framework of threshold partially HE (TPHE) and MPC&Be independent of any trusted third party against a semi-honest adversary that may compromise $m-1$
out of $m$ clients\\
\hline\hline
\multirow{4}*{\shortstack{Encryption}}
&\cite{Zhang2020Additively}&Asymmetrically split learning&Partial HE (PHE), additive noise&Achieving a lossless performance and more than $100$ times speedup\\
\cline{2-5}
&\cite{Zhang2020Additively}&Vertical tree-boosting system&HE&Revealing no information of each participant and achieving a lossless performance\\
\hline\hline
\multirow{1}*{\shortstack{Secure aggregation}}
&\cite{Tian2020FederBoost}&Vertical GBDT&Lightweight secure aggregation because the whole training relies on the order of the data instead of the
values&Achieving the same level of the area under the ROC
curve (AUC) with centralized training \\
\hline\hline
\multirow{6}*{\shortstack{Others}}
&\cite{Ang2020TIPRDC}&Privacy attributes inferring from extracted features&Adversarial training and neural network based mutual information estimator& First task-independent privacy-respecting
data crowdsourcing framework\\
\cline{2-5}
&\cite{Sharif2020Can}&Against DCM and DTWM attacks in split learning&Adding more hidden layers& Preventing privacy leakage with a slight reduction in performance\\
\cline{2-5}
&\cite{Oscar2021Label}&Against norm-based attack&Adding Gaussian noise by making the expected norm of the positive and negative gradients in a mini-batch equal&Preventing label leakage against some
extreme scenarios. \\
\hline
\end{tabular}
\label{tab:defences_knowledge_sharing}
\end{table*}
\subsection{Level 3: Sharing Results}
\begin{figure*}
\centering
    \includegraphics[width=0.95\textwidth]{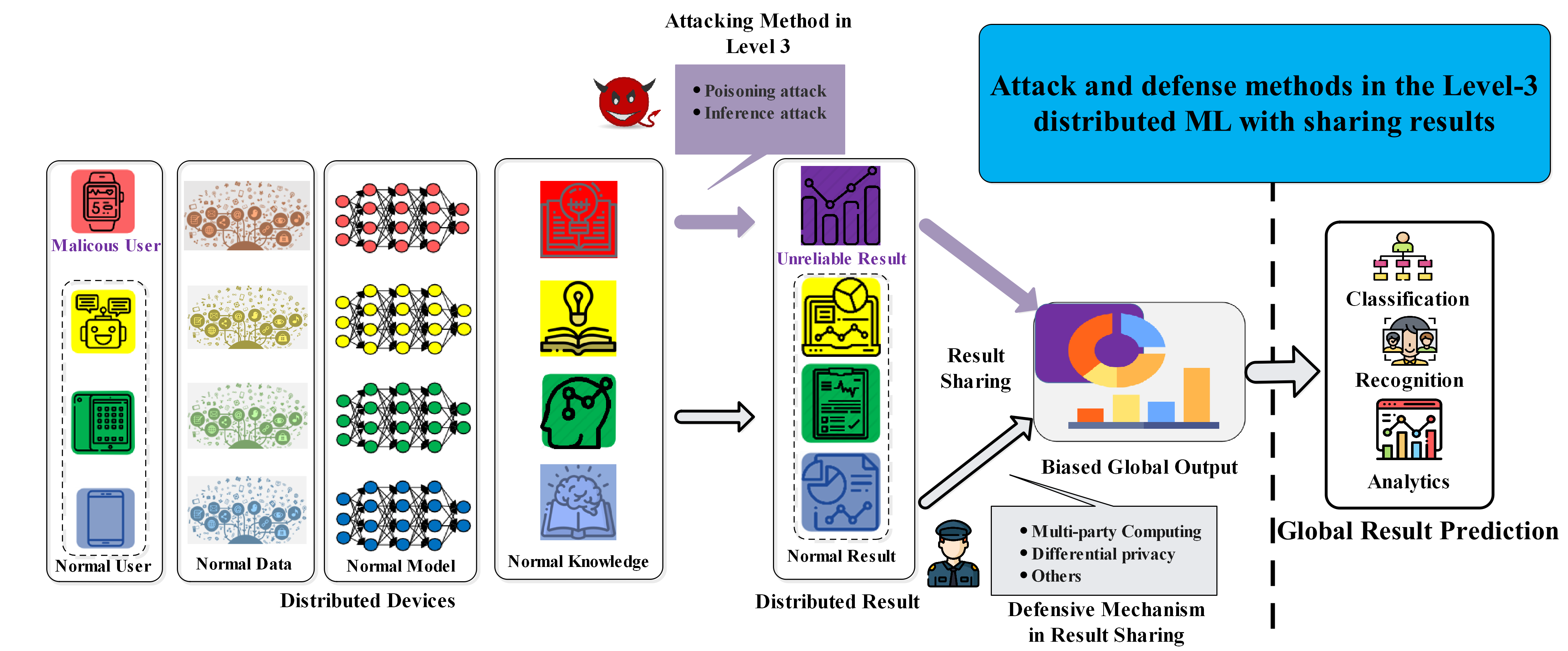}
    \caption{A breakout figure from Fig.~\ref{fig:DML}: an illustration of privacy and security issues in Level 3 distributed learning with sharing results. }
    \label{fig:result-agg}
\end{figure*}
We define the sharing results category as follows: there is no interaction or communication during the process of training. The distributed clients only share the training results after the process ends.
The history of sharing results can be traced back to ensemble ML over partitioned datasets~\cite{Trevizan2020A,breiman1996bagging}, where a number of base classifiers collectively determine the output for an instance based on a pre-defined aggregation strategy.
Ensemble techniques were originally introduced to increase the overall performance of the final classification, but it is also straightforward to utilize them for distributed ML systems~\cite{Peteiro2013A}.
The shared results~\cite{chan1993toward} in distributed learning can be either the final training models, e.g., PATE and multi-agent multi-arm bandits (MAMAB), or the prediction (output) of the models, e.g., crowd-sourcing.
\subsubsection{Threat Models}
For the result sharing models, malicious participants may exist and provide false advice or results to hinder the learning performance of other participants or the global model.
In addition, curious participants can infer some confidential information from the shared results.
\subsubsection{Taxonomy of Attacks}
As stated by da Silva \emph{et al.}\cite{Felipe2017Simultaneously}, the existence of malicious participants is a key concern in agent advises.
The work in~\cite{Umar2011A} has proposed the attack model that some of these agents might become self-interested and try to maximize car owners' utility by sending out false information.
Based on \cite{Umar2011A}, Hayes \emph{et al.}~\cite{Jamie2017LOGAN} have investigated attacks in the setting where the adversary is only permitted to access the shared results (such as the generated samples set in GAN), by retraining a local copy of the victim model.
In addition, Hilprecht \emph{et al.}~\cite{enjamin2019Monte} have proposed to count the number of generated samples that are inside an $\epsilon$-ball of the query, based on an elaborate design of distance metric.
The work in~\cite{Dingfan2020GAN} has presented the first taxonomy of membership inference attacks and focused on membership inference attack against deep generative models that reveals information about the training data used for victim models.
In spirit to Hilprecht \emph{et al.}~\cite{enjamin2019Monte}, this work scored each query by the reconstruction error directly, which does not introduce additional hyper-parameter while achieving superior performance.
We further summarize these attacks in Table~\ref{tab:attacks_result_sharing}.

\begin{table*}[hbt]
\caption{Taxonomy of attacks in Level-3 distributed ML with sharing results.}
\centering
\begin{tabular}{|m{2.5cm}<{\centering}||m{1.5cm}<{\centering}|m{3.5cm}<{\centering}|m{4cm}<{\centering}|m{4cm}<{\centering}|}
\hline
\textbf{Method}& \textbf{Ref.}& \textbf{Attacker's knowledge}& \textbf{Learning Model}& \textbf{Effectiveness}\\
\hline\hline
\multirow{1}*{\shortstack{Poisoning attack}}
&\cite{Umar2011A}&Black box& Street random waypoint (STRAW) mobility&Average
speed of vehicles in the network decreases as the percentage of
liars increases\\
\hline
\multirow{6}*{\shortstack{Inference attack}}
&\cite{Jamie2017LOGAN}&White-box, black-box &GAN& Achieving $100\%$ and $80\%$ successful at membership inferring in white-box and black-box settings, respectively\\
\cline{2-5}
&\cite{enjamin2019Monte}&Black-box&GAN, variational autoencoders (VAEs)&Success rates superior to previous work with mild assumptions\\
\cline{2-5}
&\cite{Dingfan2020GAN}&White-box, partial black-box, black-box&GAN&Consistently outperforms the state-of-the-art
models with an increasing number of generated samples\\
\hline
\end{tabular}
\label{tab:attacks_result_sharing}
\end{table*}
\begin{table*}[hbt]
\caption{Taxonomy of defenses in Level-3 distributed ML with sharing results.}
\centering
\begin{tabular}{|m{2cm}<{\centering}||m{1.5cm}<{\centering}|m{3.5cm}<{\centering}|m{4cm}<{\centering}|m{4cm}<{\centering}|}
\hline
\textbf{Method}& \textbf{Ref.}& \textbf{Use case}& \textbf{Key idea}& \textbf{Effectiveness}\\
\hline\hline
\multirow{4}*{\shortstack{DP}}
&\cite{Ye2020Differentially}& Malicious agent advising&Laplace mechanism&Reducing the impact of malicious agents without identifying them\\
\cline{2-5}
&\cite{Ren2021Multi}&Against inference attacks from any party or eavesdropper&Laplace mechanism, Bernoulli mechanism&Providing regret upper and lower bounds for MAB with local DP\\
\hline\hline
\multirow{1}*{\shortstack{MPC}}
&\cite{Zhao2018Distributed}&PATE&Training non-sensitive and unlabeled data, Securely combining the outputs by MPC& Guarantee data security\\
\hline\hline
\multirow{1}*{\shortstack{Others}}
&\cite{papernot2016semi}&PATE&The student is linked to the teachers only by their prediction capabilities and trained by ``querying the teachers about unlabelled examples''&Achieving much lower privacy budget than traditional DP approaches\\
\hline
\end{tabular}
\label{tab:defences_result_sharing}.
\end{table*}
\subsubsection{Taxonomy of Defences}
In results sharing paradigms, Tab.~\ref{tab:defences_result_sharing} summarizes the use case, key idea, and effectiveness for existing attacks. Moreover, we will discuss various defenses in model sharing frameworks as follows.
\begin{itemize}
\item[$\bullet$] \textbf{DP.} The work in~\cite{Ye2020Differentially} has proposed a novel differentially private agent advising approach, which employs the Laplace mechanism to add noise to the rewards used by student agents to select teacher agents. By using the advising approach and the DP technique, this approach can reduce the impact of malicious
agents without identifying them and naturally control communication overhead.
The work in~\cite{Ren2021Multi} adopted DP and studied the regret upper and lower bounds for MAB algorithms with a given local DP guarantee.
The differentially private PATE framework has been proposed to achieve individual privacy guarantees with provable privacy bounds~\cite{Christopher2021Personalized,Long2019Scalable}.
\item[$\bullet$] \textbf{MPC.}
Zhao~\cite{Zhao2018Distributed} has proposed to use the teacher-student framework in a more general distributed learning setting. The goal of this work is to address distributed deep learning under DP using the teacher-student paradigm. In the setting, there are a number of distributed entities and one aggregator. Each distributed entity leverages deep learning to train a teacher network on sensitive and labeled training data. The knowledge of the teacher networks is transferred to the student network at the aggregator in a privacy-preserving manner that protects the sensitive data. This transfer results from training non-sensitive and unlabeled data, which also applies secure MPC to securely combine the outputs of local ML for updating.
\item[$\bullet$] \textbf{Others.}
If an ensemble contains enough models, and each model is trained with disjoint subsets of the training data in a distributed manner, then ``any predictions made by most of the models should not be based on any particular part of the training data''~\cite{abadi2017protection}. The private aggregation of teacher ensembles (PATE) is based on this idea~\cite{papernot2016semi}.
In more detail, the ensemble is seen as a set of ``teachers'' for a new ``student'' model. The student is linked to the teachers only by their prediction capabilities, and is trained by ``querying the teachers about unlabelled examples''. The prediction result is disjointed from the training data through this process. Therefore data privacy can be protected. The privacy budget for PATE is much lower than traditional DP-based ML approaches. But it may not work in many practical scenarios as it relies on an unlabelled public dataset.
\end{itemize}

\subsubsection{Real Examples for Level-3 Distributed ML}
\begin{itemize}
    \item Large-scale online taxicab platforms, such as Uber and DiDi, have revolutionized the way people travel and socialize in cities worldwide and are increasingly becoming essential components of the modern transit infrastructure~\cite{Tang2021Value,Wang2019Adaptive}. The reinforcement learning-based dynamic bipartite graph matching approach has been adopted to assign each worker with one or more tasks to maximize the overall revenue of the platform, where the workers are dynamic while the tasks arrive sequentially. Specifically, for each worker-task pair, the platform can obtain a reward based on value-based reinforcement learning. Then, via some solutions to bipartite graph matching, such as greedy search, the platform can make near-optimal decisions. However, if the platform can obtain all workers¡¯ information and its purpose is only aiming to maximize the overall revenue, workers may be out of control. Thus, using DP to achieve fairness may be a solution~\cite{Zhu2022More}.
\end{itemize}

\subsubsection{Brief Summary}
As shown in Fig.~\ref{fig:result-agg}, although the results from ML systems are various from the raw data, they are also existing risks of privacy leakage, such as the generated samples from the generator in GAN.
Hence, several defensive mechanisms are utilized for preventing privacy leakage and against malicious participants.
\subsection{Relationship among the privacy and security issues in the four levels of distributed ML}
From level 0 to level 3, there is no certain law for the privacy and security level, but we may conclude that the forms of data show expose different degrees of information in the considered four levels. For example, compared to the prediction results in level 3, much more information can be extracted from the raw or original data in level 0. Regarding to the protection methods, designing a general mechanism for the four levels is an un-trivial task. For example, the DP-based mechanisms can be well adopted in level 0 (i.e., local DP \cite{duchi2013local,erlingsson2014rappor}), level 1 (i.e., DP in deep learning \cite{abadi2016deep}) and level 3 (i.e., PATE-GAN \cite{papernot2016semi}), but it may lose the effectiveness in level 2 (sharing knowledge).

\section{Lessons Learned}
In this section, we summarize the key lessons learned from this survey, which provides an overall view of the current research on security and privacy issues in distributed learning.
\subsection{Lessons Learned from Definitions of Security and Privacy}
The public often mixes up the terminologies of  ``Privacy'' and ``Security'', which are in fact distinctively different.
From the expression of privacy and security in distributed learning, we can learn lessons as follows.
\subsubsection{Difference between Security and Privacy}
The concerns of security and privacy issues are different~\cite{Fung2010Privacy,Cheng2020Federated,ma2021federated}. On the one hand, security issues refer to unauthorized/malicious access, change, or denial of data or learning models. Such attacks are usually launched by adversaries with expert/full knowledge of the target system. Hence, the fundamental three goals of security are confidentiality, integrity, and availability~\cite{fang2020local}. On the other hand, privacy issues generally refer to the unintentional disclosure of personal information. For example, from a side-by-side comparison of a vote registration dataset and an anonymous set of health-care sensor records (e.g., no individuals name and ID), an adversary may have the ability to identify particular individuals and the health conditions of these individuals leaks~\cite{Luo2021Feature,melis2018inference,Hu2022Membership}. This is because attributes such as gender, birth date, and zip code are the same in both datasets.
\subsubsection{Connection between Security and Privacy}
Security and privacy go hand-in-hand. Privacy issues can further induce security issues in some scenarios. If an adversary steals the private information of individuals, substantial profit from the information can be easily obtained. For example, when the adversary extracts the health conditions of an important person, he/she can blackmail the victim person by threatening to reveal the information. We know that one can envision an environment that is secure but does not guarantee privacy. Similarly, one can imagine an environment that is private, but it does not guarantee security from outsiders. Security can be achieved without privacy, but privacy cannot be achieved without security. This is because whether the security is weak or vulnerable, it will automatically affect privacy.

\subsection{Lessons Learned from Evaluations of Security and Privacy}

The evaluations on security and privacy guide the research directions in this area. In the following, we will provide some lessons by reviewing the state-of-the-art.

\subsubsection{Bayes-based Methods}
Privacy leakage can be formalized as a Bayes optimization problem from the aspect of an adversary with different assumptions on the probability distributions of the input data and interactive messages (such as gradients and extracted features). For example, the work in~\cite{Balunovic2022Bayesian} constructed a theoretical framework that can measure the expected risk that an adversary has in the process of reconstructing an input, given the joint probability distribution of inputs and their gradients. This framework can reveal the gradient leakage level by analyzing the Bayes optimal adversary, which minimizes this risk with a specific optimization problem involving the
joint distribution. DP constitutes a strong standard for privacy guarantees for algorithms on aggregate databases~\cite{dwork2008differential,abadi2016deep,Wei2020Federated}. It is defined in terms of the application-specific concept of adjacent databases and aims to hide whether one sample exists in the database. Thus, DP is defined as the detecting probability of outputs of any two adjacent databases.
\subsubsection{Experiment-based Methods}
Attack algorithms can evaluate the security and privacy levels directly. In order to evaluate the adversarial robustness of image classification tasks, large-scale experiments have been conducted and the performance of different defense methods can be evaluated~\cite {Dong2020Benchmarking}. In addition, we can apply adversaries to DP-SGD, which allows for evaluating the gap between the private information that an attacker leaks (a lower bound) and what the privacy analysis establishes as being the maximum leak (an upper bound)~\cite{Milad2021Adversary}. We can notice that attack methods constantly emerge to face advanced defense methods. Thus, the experiment-based methods need to consume a lot of computation resources, such as $3, 000$ GPU hours with parallelized over $24$ GPUs as shown in~\cite{Milad2021Adversary}.
\subsection{Lessons Learned from Attacks and Defenses}
The research on attacks and defenses in distributed learning is faced with an ``arms race'', i.e., a defense method proposed to prevent the existing attacks will be soon evaded by new attacks, and vice versa.
\subsubsection{Attacks in Distributed Learning}
Attack algorithms in the white-box scenario draw a lot of attention in the last few decades, but they seem to be impractical and can only be used as an upper bound. For example, model poisoning attacks in FL can be divided into three scenarios based on various levels of background knowledge, i.e., full knowledge, partial knowledge, and no knowledge. The attack performance decreases drastically as the background knowledge decreases~\cite{fang2020local,Xie2020DBA,Wei2022Covert}. In this context, practical attack algorithms with no knowledge should be studied to explore potential privacy and security risks. In addition, the organizer usually obtains more background knowledge than the rest of the participants. In order to mitigate the risk of the organizer being an adversary/eavesdropper, the decentralized framework can be adopted as a solution.

For the same attack purpose, different levels of distributed learning require different background knowledge, since the level of distributed learning determines interactive messages which usually contain the private information of participants, such as extracted features and neural network gradients of private data. Thus, various attack methods have emerged to infer private information or poison training process instead of unified attack schemes. For example, MIA in level 1 (sharing model) needs shadow datasets to train shadow models and then estimates the confidence of the training models~\cite{Shokri2017Membership}. We know that the shadow datasets and their distribution affect the attack performance obviously. However, how to obtain the shadow datasets becomes controversial, such as generative networks, stealing, and so on.
\subsubsection{Defenses in Distributed Learning}
Although distributed learning can achieve privacy-enhanced and scalable data sharing, it also presents some security and privacy risks. Four-level distributed learning frameworks show various risk levels of privacy leakage, due to the different interactive messages~\cite{Zhong2022Understanding,Liu2021EncoderMI}. The interactive messages usually contain the private information of participant users, such as extracted features and neural network gradients of private data. This data process can protect private data to some degree. Thus, it is of interest to study the potential privacy protection levels owing to these data process functions, and then design effective protection schemes to achieve a better trade-off between training performance and privacy.

Privacy/confidential computing for distributed learning is a high requirement compared with conventional privacy protection. However, existing privacy computing techniques usually cannot provide systematic privacy preservation, which will degrade the learning performance or training efficiency~\cite{Balcan2012Distributed,Mo2022SoK}. In addition, the protection effectiveness of different privacy computing techniques varies. For example, DP is seen as an effective method to prevent membership inference attacks by perturbing the impact on whether one instance exists in the training process. Thus, the sensitivity of interactive messages in distributed learning for DP should be carefully investigated when estimating the privacy budget. MPC is another widely used privacy computing technique. However, the transfer ability of MPC is limited and the MPC protocols for different paradigms of distributed learning need to be well-designed. Overall, it is crucial to combine these privacy computing techniques and design a general privacy-preserving framework for different paradigms of distributed learning~\cite{Bao2022Skellam,Ruan2022Private}.
\subsection{Lessons Learned from Federated Learning}
Reviewing the state-of-the-art in the field, we find that FL plays an increasingly important role in facilitating training ML models for distributed data, as highlighted as follows.
\subsubsection{The Advantages of Federated Learning}
Three classic paradigms in FL, i.e., horizontal FL, vertical FL, and federated transfer learning, can be categorized as level 1, level 2 and level 3 of distributed learning, and have the capability to address most of the challenges of training ML models in distributed scenarios. FL is an efficient approach for federated data sharing among multiple clients, in which raw data are kept on the client side, which in turn protects data privacy for tensor mining. The primary purpose of FL is to train a satisfied ML model without exposing participants' data privacy. Thus, when we select or design a training framework, both participants¡¯ data characteristics and privacy requirements should be considered. In addition, an increasing number of  advanced paradigms have emerged to handle various challenges in FL training, such as multi-modal FL~\cite{Chen2022Towards,Zong2021FedCMR,Xiong2022A}, federated knowledge distillation~\cite{Wu2022Communication,Gong2021Ensemble,Zhang2022Fine}, quantized FL~\cite{Reisizadeh2020FedPAQ} and so on, which help to construct a secure and efficient federated AI ecosystem.
\subsubsection{The Disadvantages of Federated Learning}
Although FL can benefit data privacy, security and privacy risks induced by the interactive messages also exist. Particularly, FL can be combined with other privacy techniques, such as DP, MPC, HE, and so on, to improve the privacy of local updates, by integrating them into gradient descent training to enable privacy-enhancing FL. Moreover, the security of FL-based data sharing can be improved by combining it with blockchain technology~\cite{Li2022Blockchain,Nguyen2022Latency,Deng2022Blockchain,Cui2022A}. In this context, the information of trained parameters can be appended into immutable blocks on a blockchain during client-server communications. Further, the vast communication cost in vertical FL should be noticed~\cite{Zhang2021AsySQN,Fu2021VF2Boost,Wei2022Vertical}. Specifically, in vertical FL, the total computation and communication cost is proportional to the training dataset size. In other words, the widely adopted batch computation method in horizontal FL cannot be applied to vertical FL. When facing a massive amount of data, e.g., billions of advertising data, communication, and local computation may be in many orders of magnitude, and the system may lose vitality due to limited resources, such as hardware capacity, bandwidth, and power.
\section{Research Challenges and Future Directions}
As discussed in the above sections, distributed learning systems can alleviate security and privacy concerns by advancing defense mechanisms.
In Section~\uppercase\expandafter{\romannumeral7}, we provide and reveal several critical research challenges for further improvement in system implementation.
In addition, related possible solutions are also discussed.

\begin{table*}[hbt]
\caption{Summary of challenges along with their descriptions, and possible solutions.}
\begin{tabular}{|c|c|c|}
\hline
\textbf{Challenges}                                                                                                        & \textbf{Description}                                                                                                                                                                                                                                       & \textbf{Solution}                                                                                                                       \\ \hline \hline
\multicolumn{1}{|l|}{\begin{tabular}[c]{@{}l@{}}Balance between ML performance \\ and Security/Privacy Level\end{tabular}} & \begin{tabular}[c]{@{}c@{}}The tradeoff between the Learning performance, \\ such as convergence, and the privacy and security\\  level should be well designed.\end{tabular}                                                                              & \begin{tabular}[c]{@{}c@{}}Dynamic parameter optimization \\ Specific/personalized protection mechanism\end{tabular}                    \\ \hline
Decentralized Paradigm                                                                                                     & \begin{tabular}[c]{@{}c@{}}In the distributed fashion, the regulations as well \\ as the incentives among multiple participants\\  should be investigated.\end{tabular}                                                                                    & \begin{tabular}[c]{@{}c@{}}Authentication and access control\\ Consensus design\\ Blockchain assisted distributed learning\end{tabular} \\ \hline
Complexity Reduction                                                                                                       & \begin{tabular}[c]{@{}c@{}}Distributed learning with a high complexity security \\ and privacy protection is sometimes impractical. How\\  to alleviate this complexity burden under a required\\ protection level still needs investigation.\end{tabular} & \begin{tabular}[c]{@{}c@{}}Lightweight encryption\\ High-efficiency secure protocol\\ Model compression\end{tabular}                    \\ \hline
\end{tabular}
\label{tab:Summary_of_challenges}
\end{table*}

\subsection{Balance between ML performance and Security/Privacy Level}
\subsubsection{Convergence analysis}
As mentioned above, DP has widely been adopted to train a distributed ML model, which will add random noise to gradients during the training process. However, a strict privacy guarantee usually requires a large noise variance injected,  so the DP-based training will lead to significant performance degradation.
Although existing works in~\cite{Agarwal2018cpSGD, Wei2020Federated} have explored the training performance of the differentially private distributed learning systems and provided some theoretical results,
these results can only bring out some intuitions and cannot enhance the learning performance directly.
Therefore, an accurate estimation of convergence performance on the differentially private ML training is beneficial to find a proper balance between utility and privacy.
\subsubsection{Dynamic parameter optimization}
In addition to the accurate estimation of convergence performance, dynamic parameter optimization is also a promising direction to balance the trade-off between utility and privacy. Because of privacy protection, the training performance caused by the original parameters has been changed.
Correspondingly, the conventional parameter optimization method for distributed ML also becomes inapplicable.
For example, the work in~\cite{Wei2020Federated} has developed the upper bound on the differential private FL and revealed that there exists an optimal number of communication rounds with a given privacy level. This discovery brings a new look at the communication round in FL and rethinks the choice of communication parameters.
The dynamic parameter optimization for differentially private ML has also been considered, which implements a dynamic privacy budget allocator over the course of training to improve model accuracy~\cite{Lei2019Differentially}.
Although existing dynamic optimization methods have already been proposed and proven to improve a number of distributed learning systems obviously, there is still a huge room for improvement.
\subsubsection{Specific/personalized protection mechanism}
The various requirements for different scenarios or different participants in distributed ML systems are also challenging, especially when the data distribution is non-independently identically distributed \cite{Nie2019A,Gu2020Providing}. Therefore, designing a specific/personal protection mechanism for the distributed ML system can bring out a better balance between utility and privacy.
The work in~\cite{Tao2020A} has considered a social network and achieved a proven DP requirement by perturbing each participant's option with a designated probability in each round. Combining sketch and DP techniques, the work in~\cite{Li2019Privacy} has proposed a novel sketch-based
framework, which compresses the transmitted messages via sketches to simultaneously achieve communication efficiency and provable privacy benefits.
These designs can obtain a satisfactory trade-off between utility and privacy, because of the deep combination of original scenarios and DP techniques.
Therefore, how to balance utility and privacy in the amount of distributed learning scenarios has not been fully explored.
\subsubsection{Private set intersection (PSI)}
PSI is an important step in distributed learning because of the feature or individual differences among multiple users. For example, in horizontal FL/SGD systems, we need to ensure that each record has the same features. Classical PSI protocols are third party-based PSI \cite{Baldwin1985Cryptographic,Hazay2008Constructions}, public-key-based PSI \cite{Meadows1986A,Freedman2004Efficient}, circuit-based PSI \cite{huang2012private} and OT-based PSI \cite{Dong2013When}. However, there is still a research gap that using PSI in distributed learning to investigate the tradeoff between the privacy level and the learning performance.
\subsection{Decentralized Paradigm}
\subsubsection{Authentication and access control}
The key question in adding security to a decentralized diagram is to increase the confidence that all parties involved in the system (agents, platforms, and users) will behave correctly, and can be achieved by authentication.
The identification of the parties can make up a system
and possibly establish a trusting environment between clients.
Cryptology is proven useful in a large number of authentication and access control scenarios, but it cannot address the problem of fully new participants.
In addition, a trust/reputation model has been proposed to determine the participating values for unknown clients, since it is hard for an agent to obtain complete knowledge about other participants~\cite{Petr2003Communication,Wang2007Formal,Huynh2009A}.
Consequently, how to design efficient identity certification mechanisms to uniquely authenticate known, and trusted users and agents in the system has drawn much attention.
\subsubsection{Consensus design}
Coordination and cooperative control of multi-client in distributed ML always attract lots of attention from various research communities, where a fundamental approach to achieving cooperative control is the consensus-based algorithm \cite{li2013brief}. Traditional consensus designs are mostly based on single and finite-time domain \cite{meng2011finite,MENG2012ITERATIVE}, where in reality, the dynamics of the system are usually complicated and non-linear. Therefore, a useful and effective consensus design with dynamic or unknown parameters is urgent in future research. For example, the time-varying resources and requirements for participating clients are key and un-trivial factors in design. In addition, the security of consensus also raises several issues recently \cite{Tai2021SGUARD}. How to protect the integrity of the consensus from inside or outside attackers and how to prevent private information leakage from the published consensus are other interesting research directions.
\subsubsection{Blockchain assisted distributed learning}
The reasons for implementing blockchain in a distributed learning system are to increase the interaction efficiency between participants by providing more trusted information exchange, reaching a consensus in trust conditions, assessing participant productivity or detecting performance problems, identifying intruders, allocating plans and tasks, and deploying distributed solutions and joint missions \cite{Nguyen2021Federated,Awan2019Poster}.
However, the challenges consist of assessing feasibility and finding an architectural approach for combining blockchain-based consensus algorithms with real-time distributed learning systems, while assuring incentive information exchange and compatibility with the already existent local processing protocols \cite{ma2021federated}. In addition, the incentive mechanism is also vital for the consensus design \cite{Zhan2020An,Sim2020Collaborative}.
\subsubsection{Fairness}
Fairness attracts increasing attention in recent years, especially in the scenario where multiple participants are evolved in one learning task \cite{Li2021Ditto}. A max-min fairness distributed learning system has been developed in \cite{Bistritz2020My}, where multiple clients are matched with the bandits with the minimum regret. Furthermore, collaborative fairness in FL has been investigated in \cite{Lyu2020Collaborative}. Although several works throw out the idea of fairness, there is a lack of a common definition of fairness in distributed learning. Whether attending the same rounds of training or allocating training trials according to the users' capability represents fairness is still an unclear question. In addition, the relationship between fairness with security and privacy also requires further discussion.
\subsection{Complexity Reduction}
\subsubsection{Lightweight encryption}
One of the oldest and most popular techniques used in information security is cryptography, and its use to protect valuable information is usually relying on symmetric encryption and decryption algorithms such as elliptic curve cryptography (ECC), homomorphic hash function, and secret sharing technology.
A secure lightweight ECC-Based protocol, i.e., Broadcast based Secure Mobile Agent Protocol (BROSMAP)~\cite{Hasan2017Secure}, has been improved to fulfill the needs of Multi-agent based IoT Systems in general and obtained better performance than its predecessor with the same security requirements.
HE assisted MPC framework~\cite{Xu2020VerifyNet}, enabling a participant to compute functions on values while keeping the values hidden, can allow certain mathematical operations (such as aggregation) to be performed directly on ciphertexts, without prior decryption.
However, cryptography algorithms usually require complicated computation protocols and may not be achieved efficiently.
\subsubsection{High-efficiency secure protocol}
Secure protocols are designed to enable computation over data distributed between different parties so that only the result of the computation is revealed to the participants, but no other private information.
Secure protocols usually combine several efficient security and privacy techniques, e.g., MPC, DP, and HE, and need several interactions to exchange intermediate results. However, too many interactions may increase the information leakage risk, communication, and computing overhead. Besides, it is also challenging to explore generic secure protocols over remote parties, especially for complicated scenarios and various applications. To realize an efficient communication protocol in a trusted and secure environment, an alternative way is to increase the transmission rate using an intelligent reflecting surface (IRS) by smartly reconfiguring the wireless propagation environment, with the help of massive low-cost passive reflecting elements integrated on a planar surface and to enable cover communication \cite{Wu2020Towards}.
\subsubsection{Model compression}
High accuracy of large neural networks is often achieved by paying the cost of hungry memory consumption and complex computational capability, which greatly impedes the deployment and development in distributed systems \cite{Deng2020Model}. To efficiently accelerate the learning process, privacy preservation-based methods, such as compact model \cite{Liu2022Privacy,Jin2019CaRENets}, tensor decomposition \cite{Feng2020Privacy}, data quantization \cite{Zhang2020Optimal} and network sparsification \cite{Luo2021Scalable}, are recent key advances.
\subsection{Distributed ML and Futuristic Technologies}
\subsubsection{Robotics}
Distributed ML can enhance the ability to identify and control robotics with remote and distributed control or wireless connections to clouds. This scenario requires high precision control, which raises increasing security issues and vulnerability to transmission errors~\cite{Quarta2017An,Breiling2017Secure}. How to preserve the integrity of the control system and how to prevent information leakage during data transmission needs further investigation. In addition,  ethical issues related to bionic robots are hotly debated concerns~\cite{Jokinen2021Do,sharkey2012granny}.
\subsubsection{Virtual reality (VR) and augmented reality (AR)}
ML and its distributed styles can improve the quality of generated images and videos, such as GAN and diffusion models.
With the rapid development in VR and AR-based applications, private information from generated videos may lead to personal information leakage~\cite{Devon2018Ethics,Gulhane2019Security}. Adversaries can take advantage of the fake videos to analyze the unique behaviors, personal interests, and background environments of participants~\cite{maloney2020anonymity}.
\subsubsection{Distributed quantum computing}
Quantum ML operates based on quantum mechanics, taking advantage of superposition to store and process information~\cite{Roetteler2018Quantum,fisher2014quantum}. However, if information sources are from distributed clients, information leakage and inside or outside attacks may occur during data transmission. Thus, conducting the protection on distributed ML raises several challenging problems, such as identifying attackers, ensuring the integrity and availability of transmission data, and preserving privacy.
\subsubsection{Metaverse}
Metaverse seamlessly integrates the real world with the virtual one. It allows avatars to carry out rich activities, including creation, display, entertainment, social networking, and trading. Thus, it is promising to build an exciting digital world and transform a better physical scenario by exploring the Metaverse~\cite{Yang2022,Jiang2022Reliable}.  Intuitively, the breakthroughs of AI in the real world motivate people to realize the Metaverse. For example, distributed ML via integrating distributed data from Metaverse users can provide technical support for Metaverse systems to reach or exceed the level of human learning. This can significantly affect the operational efficiency and the intelligence of the Metaverse. Intelligent voice services provide technical support, such as voice recognition and communication. However, several new security and privacy challenges that can compromise the systems or divulge users' privacy raise attention in the interaction process, such as the communication between metaverse users and service providers.
\subsubsection{Digital twin}
The digital twin can fill the gap between physical systems and digital spaces. Leveraging FL to construct digital twin models of IoT devices based on their running data has been proposed in~\cite{Lu2021Communication,Dong2019Deep}. The physical security of IoT devices is critical as they can be damaged, destroyed, or even stolen by attackers. Digital twin systems also have other priorities than the traditional network/system security requirements because of their interactions with the physical components. For instance, defects in a critical product may lead to death, injuries, or environmental damage. For this reason, safety could be ranked as the top security requirement. Safety can broadly be defined as the avoidance of harm or hazard to the physical environment and infrastructure that could occur from system faults~\cite{Enis2021Digital}. Meanwhile, the possible privacy leakage from the interactions with the physical components must also be considered.

\subsubsection{Web 3.0}
Web 3.0 has attracted considerable attention due to its unique decentralized characteristics~\cite{Chen2022When}. In Web 3.0, data presents a distributed storage structure, so there will be no central node for data management, significantly reducing the service cost of managing data. Web 3.0 emphasizes the protection of users¡¯ personal data, and therefore, as a key technology to solve the data privacy problem, privacy computing is becoming the immediate need of Web 3.0 existence. Privacy computing technology can analyze and calculate data under the premise of protecting data privacy and security, which provides a strong guarantee for the efficient and safe circulation of data across industries and organizations.
\subsubsection{Generative design AI}
Generative design uses AI to come up with multiple design variations for products or parts. This leads to a faster generation of design options than would be developed through manual design, which leads to faster product development times and more creative choices to select from. For example, the meteoric rise of diffusion models has been one of the most significant developments in ML in the past several years~\cite{Huang2021A}. Although generative design AI can improve the qualities of several tasks, it also relies on massive data and may induce several security and privacy issues, especially for fake digital assets, like photos or videos, that are indistinguishable from real things.

{\color{blue}\subsection{Development of IEEE standardizations, policy, and regulations}
Privacy and security are paramount considerations in the field of distributed learning, where data is shared and processed across various decentralized nodes. To ensure a robust and trustworthy environment for distributed learning systems, several IEEE standards, policies, and regulations come into play. These guidelines help establish a solid foundation for protecting user data and maintaining the integrity of the learning process.
\subsubsection{IEEE Standards}
\begin{itemize}
    \item IEEE 1363 (Standard Specifications for Public-Key Cryptography): Encryption is vital for securing data in distributed learning. IEEE 1363 provides specifications for public-key cryptography algorithms, ensuring confidentiality and integrity of communication in distributed systems.
    \item IEEE P2089  (Standard for Privacy Impact Assessment for Internet of Things): This standard provides a framework for assessing the privacy impact of IoT systems, which often play a crucial role in distributed learning scenarios. It guides the identification of potential privacy risks and suggests mitigation strategies.
    \item  IEEE 3652.1-2020 (Guide for Architectural Framework and Application of Federated Machine Learning)\footnote{https://standards.ieee.org/standard/3652$\_$1-2020.html}: It provides a blueprint for data usage and model building across organizations and devices while meeting applicable privacy, security and regulatory requirements in FL. In detail, the description and definition; the categories and the application scenarios to which each category applies; the performance evaluation; and the associated regulatory requirements of FL are defined.
    \item IEEE P7000 series (Model Process for Addressing Ethical Concerns During System Design): Distributed learning involves ethical considerations, and this series offers a comprehensive model process to address ethical concerns throughout system design and development. It emphasizes transparency, accountability, and user consent.
\end{itemize}

\subsubsection{Policies and Regulations}
\begin{itemize}
    \item GDPR (General Data Protection Regulation)\footnote{https://ec.europa.eu/info/law/law-topic/data-protection$\_$en}: Although not an IEEE standard, GDPR is a significant regulation that affects distributed learning. It emphasizes the protection of personal data and requires explicit user consent for data processing. Organizations handling data in distributed learning must adhere to GDPR's principles to ensure user privacy.
    \item HIPAA (Health Insurance Portability and Accountability Act)\footnote{https://www.hhs.gov/hipaa/index.html}: In healthcare-related distributed learning applications, HIPAA plays a crucial role. It sets regulations for protecting the privacy and security of patient's health information, including data used in distributed learning scenarios.
    \item NIST (National Institute of Standards and Technology) Guidelines\footnote{https://csrc.nist.gov/}: While not IEEE-specific, NIST provides guidelines on security and privacy, including those applicable to distributed systems. NIST's cybersecurity framework and privacy framework offer valuable insights for building secure and privacy-preserving distributed learning systems.
    \item IEEE Code of Ethics\footnote{https://www.ieee.org/about/ieee-code-of-ethics.html}: While not a policy or regulation in the legal sense, the IEEE Code of Ethics guides professionals working in technical fields, including distributed learning. It encourages ethical behavior, respect for privacy, and responsible decision-making.
\end{itemize}}
\section{Conclusions}
As an important and emerging technology, distributed ML has the capability to leverage the incremental amount of data in UEs to the maximum extent. However, this emergence raises increased concerns about privacy and security.  In this survey, we have proposed a new framework, which divides distributed ML into four levels for the purpose of understanding privacy and security issues. Moreover, we have discussed and summarized the state-of-the-art related to these issues and revealed the particular characteristics of adversaries at each level. In addition, several research challenges and future directions have also been discussed.
\bibliographystyle{IEEEtran}
\bibliography{sec1,sec2,sec3,sec4,sec5,sec6,sec7,sec8}

\begin{IEEEbiography}[{\includegraphics[width=1in,height=1.25in,clip,keepaspectratio]{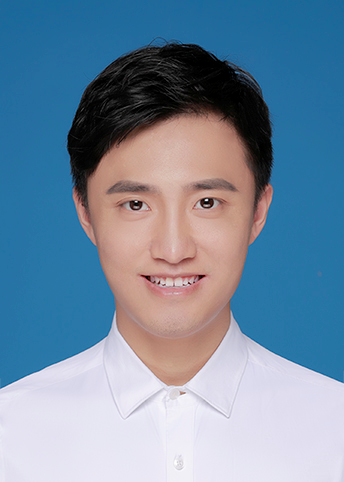}}]
{Chuan Ma} (M'19) received the B.S. degree from the Beijing University of Posts and Telecommunications, Beijing, China, in 2013 and the Ph.D. degree from the University of Sydney, Australia, in 2018. From 2018 to 2022, he worked as a lecturer at the Nanjing University of Science and Technology, and now he is a principal investigator at Zhejiang Lab, Hangzhou, China. He has published more than 40 journal and conference papers, including the best paper in WCNC 2018, and the best paper award in IEEE Signal Processing Society 2022. His research interests include stochastic geometry, wireless caching networks, and distributed machine learning, and now focuses on big data analysis and privacy-preserving.
\end{IEEEbiography}

\begin{IEEEbiography}[{\includegraphics[width=1in,height=1.25in,clip,keepaspectratio]{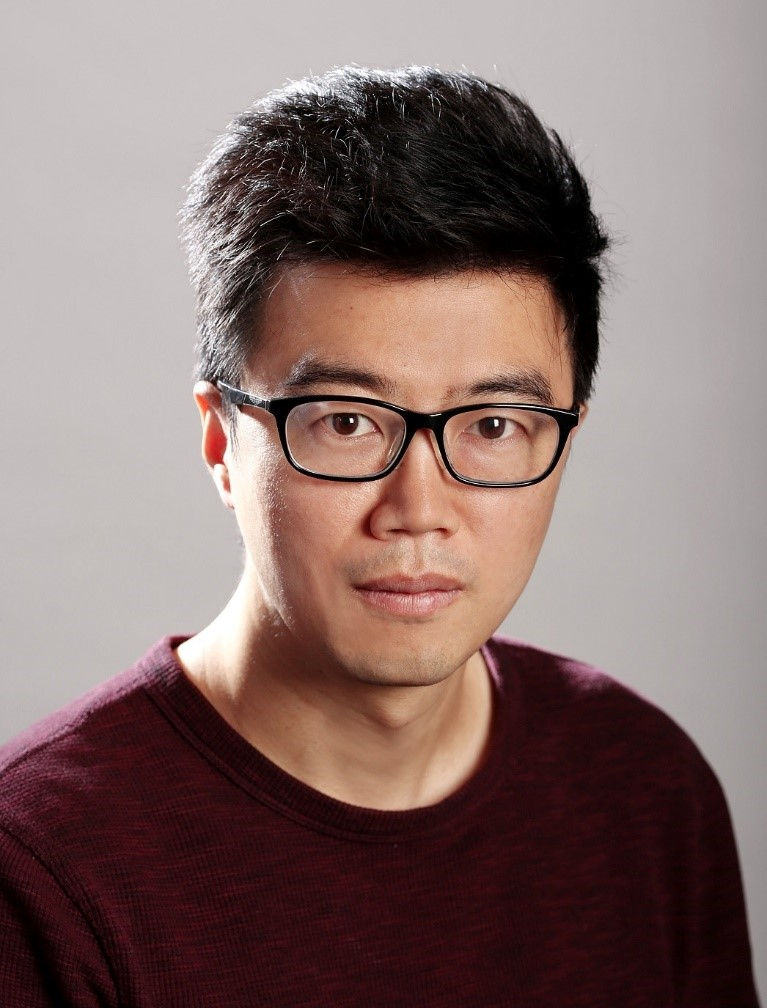}}]
{Jun Li} (SM'16) received Ph. D degree in Electronic Engineering from Shanghai Jiao Tong University, Shanghai, P. R. China in 2009. From January 2009 to June 2009, he worked in the Department of Research and Innovation, Alcatel Lucent Shanghai Bell as a Research Scientist. From June 2009 to April 2012, he was a Postdoctoral Fellow at the School of Electrical Engineering and Telecommunications, at the University of New South Wales, Australia. From April 2012 to June 2015, he was a Research Fellow at the School of Electrical Engineering, the University of Sydney, Australia. From June 2015 to now, he is a Professor at the School of Electronic and Optical Engineering, Nanjing University of Science and Technology, Nanjing, China. He was a visiting professor at Princeton University from 2018 to 2019. His research interests include network information theory, game theory, distributed intelligence, multiple agent reinforcement learning, and their applications in ultra-dense wireless networks, mobile edge computing, network privacy and security, and the industrial Internet of things. He has co-authored more than 200 papers in IEEE journals and conferences and holds 1 US patent and more than 10 Chinese patents in these areas. He is serving as an editor of IEEE Transactions on Wireless Communication and TPC member for several flagship IEEE conferences.
\end{IEEEbiography}

\begin{IEEEbiography}[{\includegraphics[width=1in,height=1.25in,clip,keepaspectratio]{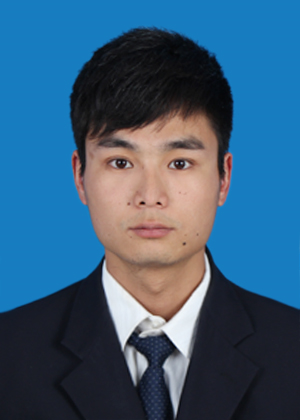}}]
{Kang Wei} (M'23) received his Ph.D. degree from Nanjing University of Science and Technology. Before that, he received the B.S. degree in information engineering from Xidian University, Xian, China, in 2014. He is currently a postdoctoral fellow at The Hong Kong Polytechnic University. He mainly focuses on privacy protection and optimization techniques for edge intelligence, including federated learning, differential privacy, and network resource allocation.
\end{IEEEbiography}

\begin{IEEEbiography}[{\includegraphics[width=1in,height=1.25in,clip,keepaspectratio]{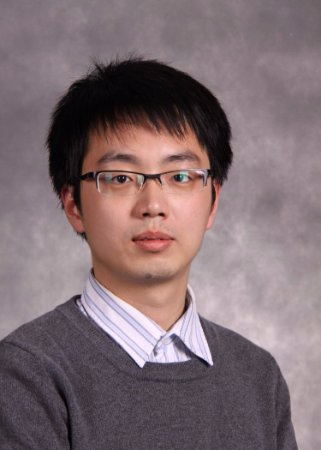}}]
{Bo Liu} received the BEng degree from the Department of Computer Science and Technology, Nanjing University of Posts and Telecommunications, Nanjing, China, in 2004. He then received the MEng. and Ph.D. Degrees from the Department of Electronic Engineering, Shanghai Jiao Tong University, Shanghai, China, in 2007 and 2010, respectively. He is currently an Associate Professor at the University of Technology Sydney, Australia. His research interests include cybersecurity and privacy, location privacy and image privacy, privacy protection, and machine learning.
\end{IEEEbiography}

\begin{IEEEbiography}[{\includegraphics[width=1in,height=1.25in,clip,keepaspectratio]{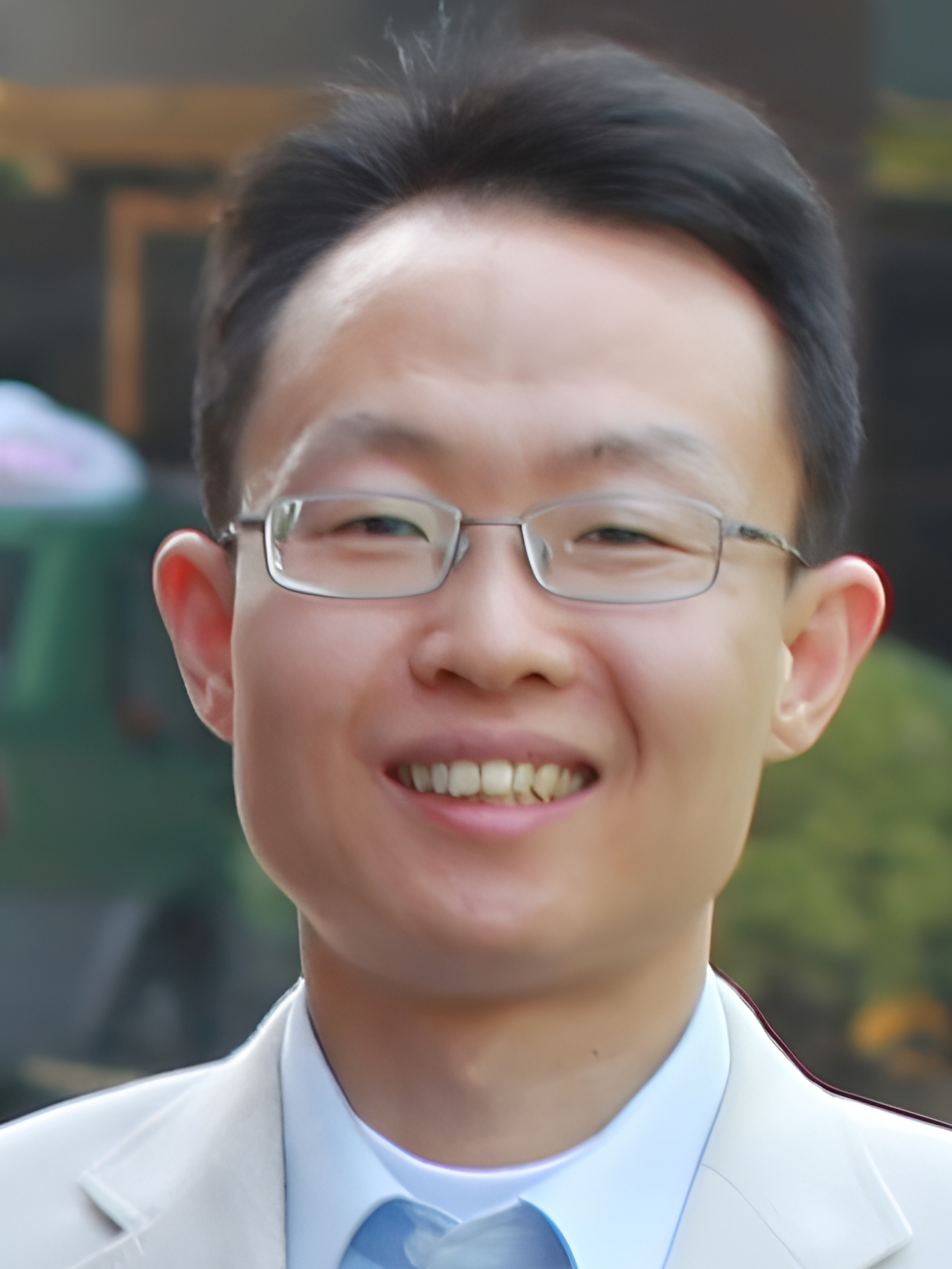}}]
{Ming Ding} (IEEE M'12-S'17) received the B.S. and M.S. degrees (with first-class Hons.) in electronics engineering from Shanghai Jiao Tong University (SJTU), Shanghai, China, and the Doctor of Philosophy (Ph.D.) degree in signal and information processing from SJTU, in 2004, 2007, and 2011, respectively. From April 2007 to September 2014, he worked at Sharp Laboratories of China in Shanghai, China as a Researcher/Senior Researcher/Principal Researcher. Currently, he is a Principal Research Scientist at Data61, CSIRO, in Sydney, NSW, Australia. His research interests include information technology, data privacy and security, and machine learning and AI. He has authored more than 200 papers in IEEE journals and conferences, all in recognized venues, and around 20 3GPP standardization contributions, as well as two books, i.e., ``Multi-point Cooperative Communication Systems: Theory and Applications'' (Springer, 2013) and ¡°Fundamentals of Ultra-Dense Wireless Networks¡± (Cambridge University Press, 2022). Also, he holds 21 US patents and has co-invented another 100+ patents on 4G/5G technologies. Currently, he is an editor of IEEE Transactions on Wireless Communications and IEEE Communications Surveys and Tutorials. Besides, he has served as a guest editor/co-chair/co-tutor/TPC member for multiple IEEE top-tier journals/conferences and received several awards for his research work and professional services, including the prestigious IEEE Signal Processing Society Best Paper Award in 2022.
\end{IEEEbiography}

\begin{IEEEbiography}[{\includegraphics[width=1in,height=1.25in,clip,keepaspectratio]{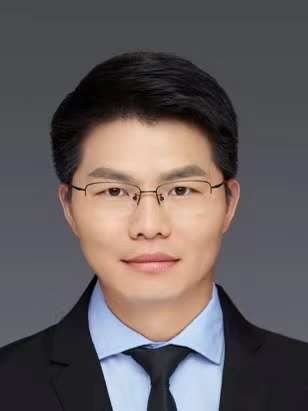}}]
{Long Yuan} is currently a professor at the School of Computer Science and Engineering, Nanjing University of Science and Technology, China. He received his Ph.D. degree from the database group of the University of New South Wales, Australia, M.S. degree and B.S. degree both from Sichuan University, China. His research interests include graph data management and analysis. He has published papers in conferences and journals including VLDB, ICDE, WWW, The VLDB Journal, and TKDE.
\end{IEEEbiography}

\begin{IEEEbiography}[{\includegraphics[width=1in,height=1.25in,clip,keepaspectratio]{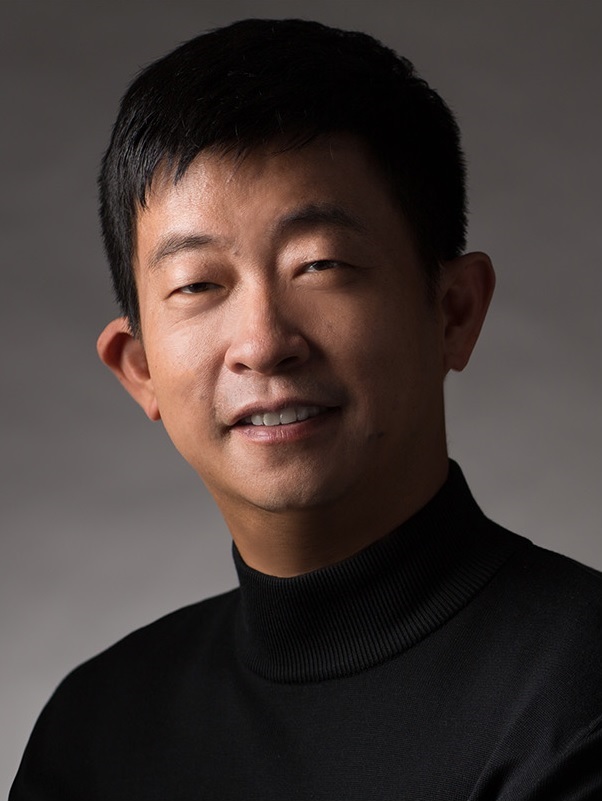}}]
{Zhu Han} (S'01¨CM'04-SM'09-F'14) received the B.S. degree in electronic engineering from Tsinghua University, in 1997, and the M.S. and Ph.D. degrees in electrical and computer engineering from the University of Maryland, College Park, in 1999 and 2003, respectively. From 2000 to 2002, he was an R$\&$D Engineer of JDSU, Germantown, Maryland. From 2003 to 2006, he was a Research Associate at the University of Maryland. From 2006 to 2008, he was an assistant professor at Boise State University, Idaho. Currently, he is a John and Rebecca Moores Professor in the Electrical and Computer Engineering Department as well as in the Computer Science Department at the University of Houston, Texas. His research interests include wireless resource allocation and management, wireless communications and networking, game theory, big data analysis, security, and smart grid.  Dr. Han received an NSF Career Award in 2010, the Fred W. Ellersick Prize of the IEEE Communication Society in 2011, the EURASIP Best Paper Award for the Journal on Advances in Signal Processing in 2015, IEEE Leonard G. Abraham Prize in the field of Communications Systems (best paper award in IEEE JSAC) in 2016, and several best paper awards in IEEE conferences. Dr. Han was an IEEE Communications Society Distinguished Lecturer from 2015-2018, AAAS fellow since 2019, and ACM distinguished Member since 2019. Dr. Han is a 1\% highly cited researcher since 2017 according to Web of Science. Dr. Han is also the winner of the 2021 IEEE Kiyo Tomiyasu Award, for outstanding early to mid-career contributions to technologies holding the promise of innovative applications, with the following citation: ``for contributions to game theory and distributed management of autonomous communication networks.''
\end{IEEEbiography}

\begin{IEEEbiography}[{\includegraphics[width=1in,height=1.25in,clip,keepaspectratio]{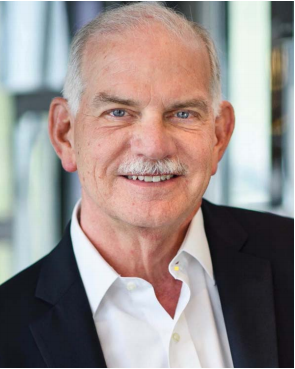}}]
{H. Vincent Poor} (S'72, M'77, SM'82, F'87) received the Ph.D. degree in EECS from Princeton University in 1977. From 1977 until 1990, he was on the faculty of the University of Illinois at Urbana-Champaign. Since 1990 he has been on the faculty at Princeton, where he is currently the Michael Henry Strater University Professor. From 2006 to 2016, he served as the dean of Princeton¡¯s School of Engineering and Applied Science. He has also held visiting appointments at several other universities, including most recently at Berkeley and Cambridge. His research interests are in the areas of information theory, machine learning, and network science, and their applications in wireless networks, energy systems, and related fields. Among his publications in these areas is the recent book \emph{Machine Learning and Wireless Communications}. (Cambridge University Press, 2022). Dr. Poor is a member of the National Academy of Engineering and the National Academy of Sciences and is a foreign member of the Chinese Academy of Sciences, the Royal Society, and other national and international academies. He received the IEEE Alexander Graham Bell Medal in 2017.
\end{IEEEbiography}

\end{document}